\documentclass[11pt]{scrartcl}

\usepackage[utf8]{inputenc}
\usepackage[T1]{fontenc}
\usepackage[english]{babel}
\usepackage{bbold}
\usepackage{lmodern}
\usepackage{microtype}
\usepackage{comment}
\usepackage{orcidlink}

\usepackage{amsmath,amssymb,amsfonts}
\usepackage{bm}
\usepackage{nicematrix}

\usepackage{graphicx}
\usepackage{hyperref}

\usepackage{geometry}
\usepackage{float}
\usepackage{xcolor}

\usepackage[format=plain,
            singlelinecheck=false,
            indention=0em,
            labelfont=bf]{caption}

\usepackage[symbol]{footmisc}

\newcommand{\email}[1]{\footnote{\texttt{#1}}}

\RedeclareSectionCommand[
  beforeskip=1.2em,
  afterskip=0.6em,
  font=\large\bfseries
]{section}

\RedeclareSectionCommand[
  beforeskip=1.0em,
  afterskip=0.4em,
  font=\normalsize\bfseries
]{subsection}

\RedeclareSectionCommand[
  beforeskip=0.8em,
  afterskip=0.3em,
  font=\normalsize\itshape
]{subsubsection}

\begin{document}

\begin{center}
    {\Large \textbf{Neutron stars in Poincaré gauge gravity with quadratic torsion} \\[1.2em]}
    {Chaitanya Vashistha$^{1, }$\email{vashisthachaitanya@gmail.com}
    \orcidlink{0009-0002-7699-389X},
    {Radouane Gannouji$^{2, }$\email{radouane.gannouji@pucv.cl}} \orcidlink{0000-0003-0749-7593},
    {Apratim Ganguly$^{1, }$\email{apratim@iucaa.in}} \orcidlink{0000-0001-7394-0755} \\[0.9em]}
{\small $^{1}$ Inter-University Centre for Astronomy and Astrophysics, Post Bag 4, Ganeshkhind, Pune 411007, India.\\
$^{2}$Instituto de Física, Pontificia Universidad Católica de Valparaíso, Av. Brasil 2950, Valparaíso, Chile.}
\end{center}

\vspace{1em}

\begin{abstract}
We study static neutron stars in an algebraic sector of Poincar\'e gauge gravity with parity-even and parity-odd quadratic torsion invariants. Since torsion is non-propagating, the contorsion equation is algebraic and can be solved in terms of the spin current. For a Weyssenhoff fluid satisfying the Frenkel condition, the metric field equations reduce to ordinary Riemannian Einstein equations sourced by an effective fluid containing spin-squared corrections. We derive the effective energy density, radial pressure, and tangential pressure, allowing both isotropic and anisotropic spin correlations. In contrast with Einstein--Cartan theory, the coefficient of the effective spin-spin interaction is not fixed, but depends on the dimensionless quadratic-torsion couplings. In the Einstein--Cartan limit, using the metric definition of the stress-energy tensor, the unpolarized spin contribution gives $w_{\mathrm{spin}}=-1/3$. We then derive the corresponding modified Tolman--Oppenheimer--Volkoff equations and solve them numerically using the DD2 equation of state. For the positive effective spin-spin coupling branch considered here, the torsion correction makes the stellar configurations more compact, lowers the maximum mass, and reduces the binding energy relative to the general-relativistic sequence. For the smooth weak-polarization profiles considered, spin-correlation anisotropy has only a negligible effect on the mass--radius relation.
\end{abstract}

\section{Introduction}

Neutron stars provide one of the most stringent astrophysical arenas for testing gravitational physics beyond general relativity (GR). Their large compactness and densities above nuclear saturation density make them sensitive not only to the equation of state of dense matter, but also to possible corrections to the gravitational dynamics. In standard relativistic stellar structure, the equilibrium of a static and spherical star is governed by the Tolman--Oppenheimer--Volkoff (TOV) equations, obtained from the Einstein equations coupled to a perfect fluid \cite{Tolman:1939jz,Oppenheimer:1939ne}. Any modification of the gravitational sector, or of the effective matter source, can therefore leave observable imprints on the mass--radius relation, the maximum supported mass, and the binding energy of compact stars \cite{Berti:2015itd,Lattimer:2000nx,Lattimer:2006xb}.

A natural extension of general relativity is obtained by allowing the spacetime connection to possess torsion. In Riemann--Cartan geometry, the metric and the connection are independent geometric objects, and torsion is associated with the antisymmetric part of the connection. The simplest realization is Einstein--Cartan theory, where torsion is not dynamical but is instead algebraically sourced by the intrinsic spin density of matter \cite{Kibble:1961ba,Hehl:1976kj,Trautman:2006fp,Shapiro:2001rz,Hammond:2002rm}. For ordinary matter this produces spin--spin contact interactions after torsion is eliminated. These corrections are quadratic in the microscopic spin density. Since the intrinsic spin carried by each particle is of order \(\hbar\), the corresponding spin-density scalar scales as
\begin{align}
q\sim \hbar^2 n^2 ,
\end{align}
where \(n\) is the particle number density. This \(\hbar^2\) suppression is the reason why spin--torsion corrections in Einstein--Cartan theory are usually expected to be irrelevant for stellar physics. In particular, Hehl et al. emphasized that, in the original Einstein--Cartan or \(U_4\) theory, such corrections are of no practical significance for stars, even for neutron stars \cite{Hehl:1976kj}. Spin fluids of Weyssenhoff type provide a phenomenological macroscopic description of matter with intrinsic spin \cite{Weyssenhoff:1947iua,Weyssenhoff:1947vye,Obukhov:1987yu,Boehmer:2006gd}.

This conclusion relies on the fact that, in Einstein--Cartan theory, the coupling between torsion and spin is fixed, so that the coefficient of the induced spin--spin interaction is not an independent parameter. A broader framework is provided by Poincaré gauge gravity, where the gravitational action may contain additional torsion and curvature invariants compatible with local Lorentz symmetry \cite{Blagojevic:2002du,Blagojevic:2013xpa,Obukhov:2018bmf,Baekler:2011jt,Pagani:2015ema,Hehl:2013qga}. In the present work we focus on the algebraic-torsion sector, in which no kinetic term for torsion is included. The leading corrections to the Einstein--Hilbert action are then the parity-even and parity-odd quadratic torsion invariants. Since torsion has mass dimension one, these \(T^2\) operators have the same mass dimension as the Ricci scalar and are controlled by dimensionless couplings. The Holst density can also be traded, up to a boundary term, for parity-odd torsion invariants through the Nieh--Yan identity \cite{Nieh:1981ww,Nieh:2007zz}.

After eliminating the non-propagating contorsion, the theory reduces to an effective Riemannian description in which the spin density generates additional spin-squared contributions to the metric field equations. Unlike in Einstein--Cartan theory, the coefficient of this effective spin--spin interaction is not fixed to its standard value: it depends on the dimensionless quadratic-torsion couplings. Thus the microscopic \(\hbar^2\) suppression remains present, but its effect can be compensated phenomenologically by an enhanced effective spin--spin coupling. This makes it possible to explore both the sign and the magnitude of the spin-induced correction in a controlled phenomenological way. This point is especially relevant for compact stars, where the density dependence of the spin correction can modify the high-density part of the stellar sequence.

In this paper we study static neutron-star configurations in Poincaré gauge gravity with algebraic quadratic torsion, coupled to a Weyssenhoff spin fluid. We first derive the metric and contorsion field equations and solve the algebraic contorsion equation in terms of the irreducible components of the spin current. We then specialize to a Weyssenhoff fluid satisfying the Frenkel condition and obtain the effective energy density, radial pressure, and tangential pressure induced by the spin sector. The isotropic spin distribution gives a perfect-fluid correction, while anisotropic spin correlations generate an effective pressure anisotropy.

After algebraic torsion is eliminated, the split between the matter energy--momentum tensor and the torsion-induced spin-squared contribution is definition-dependent. This is a general feature of the Riemannian rewriting of torsion theories. We analyze it explicitly in the Einstein--Cartan limit because that case allows a direct comparison with the Weyssenhoff-fluid literature, where the spin sector is commonly interpreted as a stiff component, \(w_{\mathrm{spin}}=1\) \cite{Obukhov:1987yu,Brechet:2007cj, Jockel:2024fps}. With the metric
definition of the stress-energy tensor used in this work, the unpolarized Einstein--Cartan limit instead gives \(w_{\mathrm{spin}}=-1/3\).

Finally, we derive the modified Tolman--Oppenheimer--Volkoff equations sourced by the effective anisotropic fluid and solve them numerically using the DD2 equation of state \cite{Typel:2009sy,Hempel:2009mc}. We analyze how the quadratic-torsion spin correction modifies the mass--radius relation, the maximum mass, and the relation between gravitational and baryonic mass. Our results show that, for the positive effective spin--spin coupling branch considered here, the torsion-induced correction makes the stellar configurations more compact and reduces the binding energy relative to the general-relativistic sequence.

\section{Poincar\'e gauge gravity with quadratic torsion}
\label{sec:pgt}

\subsection{Geometric conventions}
\label{subsec:metric-affine-conventions}

We work in a metric-compatible spacetime with torsion. The fundamental geometric objects are the metric
\(g_{\mu\nu}\) and a connection \(\Gamma^\rho_{\mu\nu}\), which is not assumed to be symmetric in its
lower indices. We follow Schouten's conventions \cite{Schouten1954RicciCalculus}, and use the Lorentzian signature $(-+++)$ with the convention $\varepsilon^{0123}=1/\sqrt{-g}$. The covariant derivative is defined by
\begin{align}
    & \nabla_\mu V^\nu=\partial_\mu V^\nu+\Gamma_{\mu\lambda}^\nu V^\lambda\,,\quad  \nabla_\mu V_\nu=\partial_\mu V_\nu-\Gamma_{\mu\nu}^\lambda V_\lambda\,
\end{align}
and the torsion is
\begin{align}
\label{def:torsion}
    & T_{\mu\nu}{}^{\sigma} = \frac{1}{2} \Bigl(\Gamma_{\mu\nu}^{\sigma}-\Gamma_{\nu\mu}^{\sigma}\Bigr)\,,
\end{align}
where $\Gamma$ denotes the full connection. The connection can be decomposed into the Levi--Civita connection $\{^{\rho}_{\mu\nu}\}$ plus an additional tensor, called the contorsion:
\begin{align}
\label{def:contorsion}
    \Gamma_{\mu\nu}^{\sigma} = \{{}_{\mu\nu}^{\sigma}\}+K_{\mu\nu}{}^\sigma\,.
\end{align}
In the usual Einstein--Cartan theory or Poincaré gauge theory, one imposes metric compatibility,
\begin{align}
\nabla_{\lambda} g_{\mu\nu}=0 ,
\end{align}
so that the connection has torsion but no nonmetricity. Using
\begin{align}
    \nabla_{\lambda} g_{\mu\nu}=\overset{\,*}{\nabla}_{\lambda} g_{\mu\nu}-K_{\lambda \mu\nu}-K_{\lambda\nu\mu}
\end{align}
where $\overset{\,*}{\nabla}$ is the covariant derivative based on the Levi-Civita connection, we obtain
\begin{align}
\label{eq:antis-cont}
K_{\lambda\mu\nu}=-K_{\lambda\nu\mu}
\end{align}
Next, using the combination $\nabla_{(\mu} g_{\nu \rho)}=0$ we find
\begin{align}
    K_{\mu\nu}{}^\sigma=T_{\mu\nu}{}^\sigma+T_{\lambda\nu}{}^\rho g_{\mu\rho} g^{\lambda\sigma}-T_{\mu\lambda}{}^\rho g_{\nu\rho} g^{\lambda\sigma} = g^{\lambda\sigma} \Bigl(T_{\mu\nu\lambda}+T_{\lambda\nu \mu} -T_{\mu\lambda \nu} \Bigr) = g^{\lambda\sigma} \Bigl(T_{\mu\nu\lambda}-T_{\nu \lambda \mu} +T_{\lambda \mu \nu} \Bigr)\label{schouten2}
\end{align}
Finally, the curvature is defined by
\begin{align}
    & R_{\nu\mu\lambda}{}^\sigma = \partial_\nu \Gamma_{\mu\lambda}^\sigma - \partial_\mu \Gamma_{\nu\lambda}^\sigma + \Gamma_{\nu\rho}^\sigma \Gamma_{\mu\lambda}^\rho - \Gamma_{\mu\rho}^\sigma \Gamma_{\nu\lambda}^\rho\,,\quad R_{\mu\nu} = R_{\sigma\mu\nu}{}^\sigma \,, \quad R=g^{\mu\nu}R_{\mu\nu}
\end{align}
Using the decomposition of the affine connection into its Levi-Civita part and the contorsion \eqref{def:contorsion}, the Ricci scalar can be written as
\begin{align}
\label{eq:Riemann}
    R=\overset{\,*}{R}+2\overset{\,*}{\nabla}_\alpha K^\alpha+g^{\mu\nu} g^{\alpha\beta} g^{\rho \sigma}\Bigl(K_{\alpha\rho\beta}K_{\mu\nu\sigma}-K_{\mu\rho\beta}K_{\alpha\nu\sigma}\Bigr)\,,\quad K^\alpha \equiv K_\mu{}^{\mu\alpha}
\end{align}
The second term is a total derivative and will not contribute to the bulk equations of motion under the
boundary conditions assumed below.

The theory considered in this work belongs to the algebraic-torsion sector of Poincar\'e gauge gravity.
The connection is independent, but the action contains no kinetic term for torsion. After the action is
rewritten in terms of \(g_{\mu\nu}\) and \(K_{\mu\nu}{}^\rho\), the contorsion appears algebraically in the
bulk Lagrangian. Its field equation can therefore be solved locally in terms of the spin density of matter.

\subsection{Algebraic quadratic-torsion action}

A simple algebraic-torsion sector of Poincar\'e gauge gravity is obtained by extending the Einstein--Cartan action with quadratic torsion invariants. From an effective-field-theory viewpoint,
these are the leading parity-even operators beyond the Einstein--Hilbert term:
the Ricci scalar \(R\) has mass dimension two, while torsion has mass dimension
one, so \(T^2\) has the same mass dimension (and derivative order) as \(R\).
The action can therefore be written as \cite{Hayashi:1979qx,Sezgin:1979zf,Sezgin:1981xs,Kuhfuss:1986rb,Pagani:2015ema,Baekler:2011jt}
\begin{align}
S
=
\frac{1}{2\kappa}
\int d^4 x \, \sqrt{-g}
\Bigl[
R + a_1 \, T_{\mu\nu\rho} T^{\mu\nu\rho}
+ a_2 \, T_{\mu\nu\rho} T^{\mu\rho\nu}
+ a_3 \, T_{\mu} T^{\mu}
\Bigr]
+
S_m \,,\quad \text{with}~~~T_\mu=T_{\mu\rho}{}^\rho
\end{align}
Since the torsion-quadratic operators have mass dimension two, the coefficients
\(a_1\), \(a_2\), and \(a_3\) are dimensionless. Consequently, they are not
suppressed by any cutoff scale and are genuine free parameters of the
low-energy theory. The choice
\begin{align}
a_1 = a_2 = a_3 = 0
\end{align}
reduces to standard Einstein-Cartan theory. Higher-dimension operators in the
gravitational sector (e.g. \(R^2\), \(RT^2\), \((\nabla T)^2\), \(T^3\), \dots)
require dimensionful coefficients and are expected to be suppressed by
appropriate powers of a cutoff scale \(\Lambda\). Since torsion remains algebraic, the coefficients \(a_1,a_2,a_3\) control how strongly the contorsion responds to the spin density. In this sense they modify the effective spin--spin interaction generated after torsion is eliminated. The Lagrangian can also be extended by adding parity-odd torsion invariants,
\begin{align}
S_{\text{odd}}
=\frac{1}{2\kappa}\int d^4x\sqrt{-g}\Bigl[\frac{1}{\gamma}\varepsilon^{\mu\nu\rho\sigma}R_{\mu\nu\rho\sigma}+b_1\varepsilon^{\mu\nu\rho\sigma}T_{\mu\nu}{}^{\lambda}T_{\rho\sigma\lambda}
+ b_2\varepsilon^{\mu\nu\rho\sigma}T_{\mu\nu\rho}T_{\sigma}\Bigr]
\end{align}
The first term corresponds to the Holst density with Barbero--Immirzi parameter $\gamma$, while the torsion terms are written with parameters $(b_1,b_2)$, even though the historical convention uses $(d_1,d_4)$. 

Using the Nieh--Yan identity \cite{Nieh:1981ww,Nieh:2007zz}, the Holst density differs from a linear combination of the two parity-odd torsion invariants by a boundary term. Therefore, modulo boundary terms, its effect can be absorbed into a redefinition of \(b_1\) and \(b_2\), see \hyperref[appendixA]{Appendix~\ref*{appendixA}}. The action can then be written as
\begin{align}
S
=
\frac{1}{2\kappa}
\int d^4 x \, \sqrt{-g}
\Bigl[
R + a_1 \, T_{\mu\nu\rho} T^{\mu\nu\rho}
+ a_2 \, T_{\mu\nu\rho} T^{\mu\rho\nu}
+ a_3 \, T_{\mu} T^{\mu}
+b_1\, \varepsilon^{\mu\nu\rho\sigma}T_{\mu\nu}{}^{\lambda}T_{\rho\sigma\lambda}
+ b_2\,\varepsilon^{\mu\nu\rho\sigma}T_{\mu\nu\rho}T_{\sigma}
\Bigr]
+
S_m 
\end{align}

\subsection{Metric and contorsion field equations}

The equations of motion follow from independent variations with respect to the contorsion and the metric. For this purpose, we first rewrite the action as a functional of \(g^{\mu\nu}\) and \(K_{\mu\nu\rho}\). Notice that, in the variational calculation, we do not take $g^{\mu\nu}$ and $K_{\mu\nu}{}^\rho$ as the independent variables. Instead, we use $g^{\mu\nu}$ and $K_{\mu\nu\rho}$, since this choice is directly adapted to the symmetry property \eqref{eq:antis-cont}. The allowed variations then satisfy the same antisymmetry in the last two indices, so that the projection onto the independent components of the contorsion is immediate. In the mixed-index notation $K_{\mu\nu}{}^\rho$, the corresponding constraint is metric-dependent, and the projection has to be implemented only after lowering the upper index. This makes the variation less transparent.

 Substituting the decomposition \eqref{eq:Riemann} into the action yields
\begin{align}
    S &= \frac{1}{2\kappa}\int {d^4}x \sqrt{-g}\left[\overset{\,*}{R}
+\Bigl(\frac{a_1}{2}-\frac{a_2}{4}\Bigr)K_{\mu\nu\rho}K^{\mu\nu\rho}+\Bigl(1-\frac{a_1}{2}+\frac{3a_2}{4}\Bigr)K_{\mu\nu\rho}K^{\nu\mu\rho}+\Bigl(\frac{a_3}{4}-1\Bigr)\,K_{\lambda}{}^{\lambda}{}_{\mu}K_{\rho}{}^{\rho\mu}\right.\nonumber\\
&\left.~~~
+b_1\,\varepsilon^{\mu\nu\rho\sigma}K_{\mu\nu}{}^{\lambda}K_{\rho\sigma\lambda}
+\frac{b_2}{2}\,\varepsilon^{\mu\nu\rho\sigma}K_{\mu\nu\rho}K_{\lambda}{}^{\lambda}{}_{\sigma}
\right]+ S_m + \text{boundary terms}
\label{action_contortion}
\end{align}
Varying the action with respect to the metric yields
\begin{align}
\label{eq:Einstein}
    \overset{\,*}{R}_{\mu\nu} -\frac{1}{2} \overset{\,*}{R} g_{\mu\nu} = \kappa T_{\mu\nu}+S_{\mu\nu}\,.
\end{align}
Here
\begin{align}
    T_{\mu\nu}=-\frac{2}{\sqrt{-g}}\frac{\delta S_m}{\delta g^{\mu\nu}}\,,
\end{align}
and $S_{\mu\nu}$ denotes the torsion-induced spin correction to the metric field equations:
\begin{align}
    S_{\mu\nu} =& \left(\frac{a_1}{4} - \frac{a_2}{8}\right) g_{\mu\nu} K_{\alpha \beta \gamma} K^{\alpha \beta \gamma}
    + \left(\frac{a_2}{2} - a_1\right) K_{\alpha\nu \beta} K^{\alpha}{}_{\mu}{}^{\beta}
    + \left(\frac{1}{2} - \frac{a_1}{4} + \frac{3 a_2}{8}\right) g_{\mu\nu} K^{\alpha \beta \gamma} K_{\beta \alpha \gamma}\nonumber\\
    & - \left(1 - \frac{a_1}{2} + \frac{3 a_2}{4}\right) \left( K^{\alpha}{}_{\mu}{}^{\beta} K_{\beta \nu \alpha} + K_{\alpha\mu \beta} K_{\nu}{}^{\alpha \beta} + K_{\alpha \nu \beta} K_{\mu}{}^{\alpha \beta} \right)
    - b_1 K^{\gamma}{}_{\beta}{}^{\delta}\Big(\varepsilon_{\nu \alpha \gamma \delta} K^{\alpha}{}_{\mu}{}^{\beta} + \varepsilon_{\mu \alpha \gamma \delta} K^{\alpha}{}_{\nu}{}^{\beta}\Big)\nonumber\\
    & - b_1 \varepsilon_{\alpha \beta \gamma \delta} K^{\alpha}{}_{\mu}{}^{\beta} K^{\gamma}{}_{\nu}{}^{\delta}
    + b_1 \varepsilon_{\alpha \gamma \delta \rho} g_{\mu\nu} K^{\alpha \beta \gamma} K^{\delta}{}_{\beta}{}^{\rho} 
    + \left( \frac{a_2}{4} - \frac{a_1}{2}\right) K_{\mu}{}^{\alpha \beta} K_{\nu \alpha \beta}
    - b_1 K^{\gamma}{}_{\alpha}{}^{\delta} \left( \varepsilon_{\nu \beta \gamma \delta} K_{\mu}{}^{\alpha \beta} + \varepsilon_{\mu \beta \gamma \delta} K_{\nu}{}^{\alpha \beta} \right)\nonumber\\
    & + \left(\frac{a_3}{8}-\frac{1}{2}\right) g_{\mu\nu} K_\alpha K^\alpha
    + \frac{1}{2} b_2 \left( \varepsilon_{\nu \alpha \beta \gamma} K^{\alpha}{}_{\mu}{}^{\beta} + \varepsilon_{\mu \alpha \beta \gamma} K^{\alpha}{}_{\nu}{}^{\beta} \right) K^\gamma
    - \frac{1}{4} b_2 \left( \varepsilon_{\nu \alpha \beta \gamma} K_{\mu}{}^{\alpha \beta} + \varepsilon_{\mu \alpha \beta \gamma} K_{\nu}{}^{\alpha \beta} \right) K^\gamma \nonumber\\
    & + \left(1-\frac{a_3}{4}\right) K_{\mu} K_{\nu}-\frac{1}{2} b_2 g_{\mu\nu} K^\alpha \widetilde{K}_\alpha
    + 2 K_{(\mu\nu)}{}^{\alpha} \left(\left(1-\frac{a_3}{4}\right) K_\alpha + \frac{b_2}{4} \widetilde{K}_\alpha\right)
    + \frac{1}{2} b_2  K_{(\mu} \widetilde{K}_{\nu)}\,,
\label{torsion_contri_metric}
\end{align}
where we have defined the contorsion trace
\begin{align}
    K^\alpha \equiv K_\mu{}^{\mu\alpha}
\end{align}
and the dual trace
\begin{align}
\widetilde K^\alpha \equiv \varepsilon^{\alpha\mu\nu\rho}K_{\mu\nu\rho}\,.
\end{align}
Variation with respect to the contorsion tensor yields
\begin{align}
\label{schouten1}
&\left(a_1-\frac{a_2}{2}\right)\,K^{\alpha\beta\gamma}
+\Bigl(1-\frac{a_1}{2}+\frac{3a_2}{4}\Bigr)\Bigl(K^{\beta\alpha\gamma}-K^{\gamma\alpha\beta}\Bigr)
+\Bigl(\frac{a_3}{4}-1\Bigr)\Bigl(g^{\alpha\beta}K^{\gamma}-g^{\alpha\gamma}K^{\beta}\Bigr) \nonumber\\
&\quad
+b_1\Bigl(\varepsilon^{\alpha\beta\mu\nu}K_{\mu\nu}{}^{\gamma}-\varepsilon^{\alpha\gamma\mu\nu}K_{\mu\nu}{}^{\beta}\Bigr)
+\frac{b_2}{2}\,\varepsilon^{\alpha\beta\gamma\mu}K_\mu
+\frac{b_2}{4}\Bigl(g^{\alpha\gamma}\widetilde K^\beta - g^{\alpha\beta}\widetilde K^\gamma\Bigr)
=2\kappa\,\tau^{\alpha\beta\gamma}\,,
\end{align}
where we have defined the spin density tensor
\begin{align}
\label{eq:contorsion}
    \tau^{\alpha\beta\gamma}\equiv -\frac{1}{\sqrt{-g}} \frac{\delta S_m}{\delta K_{\alpha\beta\gamma}}\,.
\end{align}

\subsection{Algebraic solution for the contorsion and effective Einstein equation}

The effective metric equation is obtained by eliminating the non-propagating contorsion from Eq.~\eqref{eq:Einstein} by means of the field equation \eqref{schouten1}. To solve this equation algebraically, we decompose the contorsion into its irreducible trace, axial-trace, and purely tensor parts,
\begin{align}   
K^{\alpha \beta \gamma}
=
\frac{1}{3}
\left(
g^{\alpha \beta} K^\gamma
-
g^{\alpha \gamma} K^\beta
\right)
+
\frac{1}{6}
\varepsilon^{\alpha \beta \gamma \delta}
\widetilde K_\delta
+
q^{\alpha \beta \gamma},
\label{contortion_decomp}
\end{align}
where the purely tensor part satisfies
\begin{align}
g_{\alpha\beta}q^{\alpha\beta\gamma}=0,
\qquad
q^{\alpha\beta\gamma}=-q^{\alpha\gamma\beta},
\qquad
q^{[\alpha\beta\gamma]}=0 .
\end{align}
We use the analogous irreducible decomposition for the spin density tensor,
\begin{align}
\tau^{\alpha \beta \gamma}
=
\frac{1}{3}
\left(
g^{\alpha \beta} \tau^\gamma
-
g^{\alpha \gamma} \tau^\beta
\right)
+
\frac{1}{6}
\varepsilon^{\alpha \beta \gamma \delta}
\widetilde{\tau}_\delta
+
t^{\alpha \beta \gamma},\qquad \tau^\alpha \equiv \tau_\mu{}^{\mu\alpha}\,,\qquad \widetilde\tau^\alpha 
\equiv \varepsilon^{\alpha\mu\nu\rho}\tau_{\mu\nu\rho}\,.
\end{align}
where
\begin{align}
g_{\alpha\beta}t^{\alpha\beta\gamma}=0,
\qquad
t^{\alpha\beta\gamma}=-t^{\alpha\gamma\beta},
\qquad
t^{[\alpha\beta\gamma]}=0 .
\end{align}
The trace and axial-trace components are obtained by projecting Eq.~\eqref{schouten1} onto the corresponding irreducible sectors. Taking the trace gives
\begin{align}
\label{eq:K1}
\left(\frac{a_1}{2}+\frac{a_2}{4}+\frac{3a_3}{4}-2\right)K^\gamma
+\left(b_1-\frac{3b_2}{4}\right)
\widetilde K^{\gamma}
=2\kappa\,g_{\alpha\beta}\tau^{\alpha\beta\gamma}
\equiv 2\kappa\,\tau^{\gamma}.
\end{align}
Taking the completely antisymmetric part of Eq.~\eqref{schouten1} gives
\begin{align}
\label{eq:K2}
2(a_1-a_2-1)\,\widetilde K_\sigma+(3b_2-4b_1)\,K_\sigma
=2\kappa\,\widetilde\tau_\sigma\,,
\end{align}
Solving the two linear equations (\ref{eq:K1},\ref{eq:K2}) gives
\begin{align}
K^\gamma
&=\frac{2\kappa}{\Delta}\Bigl[8(a_1-a_2-1)\,\tau^\gamma
+\left(3b_2-4b_1\right)\widetilde\tau^\gamma\Bigr],
\\
\widetilde K^\gamma
&=\frac{2\kappa}{\Delta}\Bigl[4\left(4b_1-3 b_2\right)\,\tau^\gamma
+\left(2 a_1+a_2+3 a_3-8\right)\widetilde\tau^\gamma\Bigr]\,,
\end{align}
where
\begin{align}
    \Delta = 2 (a_1-a_2-1)(2a_1+a_2+3 a_3-8)+(4 b_1-3 b_2)^2\,.
\end{align}
This determines only the trace and axial-trace components of the contorsion. To determine the purely tensor part, we introduce the linear map \(J\), acting on rank-three tensors antisymmetric in their last two indices, by
\begin{align}
(JX)^{\alpha\beta\gamma}
\equiv
\varepsilon^{\alpha\beta\mu\nu}X_{\mu\nu}{}^\gamma
-
\varepsilon^{\alpha\gamma\mu\nu}X_{\mu\nu}{}^\beta .
\end{align}
On the irreducible tensor subspace defined by
\begin{align}
g_{\alpha\beta}X^{\alpha\beta\gamma}=0,
\qquad
X^{\alpha\beta\gamma}=-X^{\alpha\gamma\beta},
\qquad
X^{[\alpha\beta\gamma]}=0\,,
\end{align}
one has
\begin{align}
J^2=-1 .
\end{align}
This identity follows by a direct contraction of two Levi-Civita tensors together with the irreducibility conditions above. Projecting Eq.~\eqref{schouten1} onto this tensor subspace, or equivalently
substituting the irreducible decompositions and retaining only the purely tensor
sector, gives
\begin{align}
\lambda q^{\alpha\beta\gamma}
+
b_1(Jq)^{\alpha\beta\gamma}
=
2\kappa t^{\alpha\beta\gamma},
\end{align}
where
\begin{align}
\lambda
=
\left(a_1-\frac{a_2}{2}\right)
+
\left(1-\frac{a_1}{2}+\frac{3a_2}{4}\right)
=
1+\frac{a_1}{2}+\frac{a_2}{4}.
\end{align}
Since \(J^2=-1\) on this sector, the operator
\(\lambda I+b_1J\) is inverted as
\begin{align}
(\lambda I+b_1J)^{-1}
=
\frac{\lambda I-b_1J}{\lambda^2+b_1^2},
\qquad
\lambda^2+b_1^2\neq 0.
\end{align}
Therefore
\begin{align}
q^{\alpha\beta\gamma}
=
\frac{2\kappa}{\lambda^2+b_1^2}
\left[
\lambda t^{\alpha\beta\gamma}
-
b_1(Jt)^{\alpha\beta\gamma}
\right],
\end{align}
or explicitly
\begin{align}
q^{\alpha\beta\gamma}
& =2\kappa
\frac{\left(1+\frac{a_1}{2}+\frac{a_2}{4} \right) t^{\alpha\beta\gamma}
-
b_1\left(\varepsilon^{\alpha\beta\mu\nu}t_{\mu\nu}{}^\gamma-\varepsilon^{\alpha\gamma\mu\nu}t_{\mu\nu}{}^\beta\right)}{\left(1+\frac{a_1}{2}+\frac{a_2}{4}\right)^2+b_1^2}\\
&=2\kappa
\frac{\left(1+\frac{a_1}{2}+\frac{a_2}{4} \right) t^{\alpha\beta\gamma}
-
\frac{b_1}{2}\varepsilon^{\beta\gamma \mu\nu}t^\alpha{}_{\mu\nu}}{\left(1+\frac{a_1}{2}+\frac{a_2}{4}\right)^2+b_1^2}
\end{align}
This completes the algebraic solution for the contorsion in terms of the spin density. Substituting this solution back into Eq.~\eqref{eq:Einstein} gives a purely Riemannian effective metric equation
\begin{align}
    \overset{\,*}{R}_{\mu\nu} -\frac{1}{2} \overset{\,*}{R} g_{\mu\nu} = \kappa T_{\mu\nu}+S_{\mu\nu}\,.
\end{align}
with
\begin{align}
\label{eq:fullS}
{S}_{\mu\nu}
=
\frac{\kappa^{2}}
{2\Delta \Big[(4+2a_{1}+a_{2})^{2}+16b_{1}^{2}\Big]^{2}}
\left[
\mathcal{M}_{\mu\nu}
+
g_{\mu\nu}\mathcal{T}
\right],
\end{align}
where the tensors $\mathcal{M}_{\mu\nu}$ and $\mathcal{T}$ are given in
\hyperref[appendixB]{Appendix~\ref*{appendixB}}. \footnote{The Mathematica notebook used in the derivation presented in this section is publicly available at: \\ https://github.com/ch-vashistha/Neutron-stars-Poincare-gauge-gravity}

\subsection{Einstein--Cartan limit}

The Einstein--Cartan limit is obtained by switching off all quadratic torsion
couplings in the general theory, namely
\begin{align}
a_1=a_2=a_3=b_1=b_2=0 .
\end{align}
which gives
\begin{align}
    S_{\mu\nu} =& 
    \frac{1}{2}  g_{\mu\nu} K^{\alpha \beta \gamma} K_{\beta \alpha \gamma}-\frac{1}{2} g_{\mu\nu} K_\alpha K^\alpha+ K_{\mu} K_{\nu}
    + 2 K_{(\mu\nu)}{}^{\alpha} K_\alpha- \left( K^{\alpha}{}_{\mu}{}^{\beta} K_{\beta\nu \alpha} + K_{\alpha\mu \beta} K_{\nu}{}^{\alpha\beta} + K_{\alpha\nu \beta} K_{\mu}{}^{\alpha\beta} \right) \,.
\end{align}
The contorsion is still non-propagating and is determined algebraically by the
spin density. Substituting the Einstein--Cartan solution for \(K_{\mu\nu}{}^\rho\)
into the expression above gives the spin-squared correction to the Riemannian Einstein equation
\begin{align}
S_{\mu\nu}=&\kappa^2\left(
2 \tau_{\mu}\tau_{\nu}
-2 \tau_{\alpha \beta \mu}\tau^{\alpha \beta}{}_{\nu}
-2 \tau_{\alpha \beta \mu}\tau^{\beta \alpha}{}_{\nu}
+4 \tau_{\alpha \beta (\mu}\tau_{\nu)}{}^{\alpha \beta}
- \tau_{\mu}{}^{\alpha \beta }\tau_{\nu \alpha \beta}
+4 \tau^{\alpha }\tau_{(\mu\nu) \alpha }\right.\nonumber\\
&\left.+g_{\mu\nu}\Bigl(\frac{1}{2}\tau_{\alpha \beta \gamma}\tau^{\alpha \beta \gamma}
+ \tau_{\alpha \beta \gamma}\tau^{\beta \alpha \gamma}
-\tau_{\alpha }\tau^{\alpha }\Bigr)\right)
\end{align}
The same equation is derived independently in \hyperref[appendixC]{Appendix~\ref*{appendixC}}, where a different convention is used.

\section{Weyssenhoff fluid}

\subsection{Spin current and Frenkel condition}

The Weyssenhoff fluid is a phenomenological continuum model for matter with intrinsic spin \cite{Weyssenhoff:1947iua,Weyssenhoff:1947vye,Obukhov:1987yu,deBerredo-Peixoto:2009yvf,Boehmer:2006gd}. It is not a gravitational theory by itself, but a constitutive
description of the matter currents which can be coupled, for example, to a
Poincar\'e gauge theory. The basic macroscopic variables are the fluid
four-velocity \(u^\mu\), normalized as
\begin{align}
    u^\mu u_\mu=-1 ,
\end{align}
and an antisymmetric spin
density tensor
\begin{align}
    s_{\mu\nu}=-s_{\nu\mu}.
\end{align}
The tensor \(s_{\mu\nu}\) describes the intrinsic angular momentum per unit comoving volume carried by the fluid elements. 

The defining assumption of the Weyssenhoff model is that the spin is convected
with the fluid. Thus the spin current is taken to have the form
\begin{align}
    \tau_{\alpha \mu\nu}
    =
    u_\alpha s_{\mu\nu}\,.
\end{align}
This ansatz means
that there is no independent spin flux relative to the matter flow: the spin
density is transported by the same four-velocity \(u^\mu\) that defines the
fluid motion. The model is usually supplemented by the Frenkel condition
\begin{align}
    s_{\mu\nu}u^\nu=0 .
\end{align}
This condition removes the components of $s_{\mu\nu}$ along $u^\mu$. Therefore \(s_{\mu\nu}\) contains three independent components, corresponding
to the ordinary spatial spin density. It can be decomposed covariantly in terms
of a spatial pseudovector \(s^\mu\) as
\begin{align}
\label{eq:decomp}
s_{\mu\nu}=\epsilon_{\mu\nu\rho\sigma}u^\rho s^\sigma,
\qquad
u_\mu s^\mu=0,
\end{align}
and therefore, the torsion-induced contribution to the Einstein equation \eqref{eq:fullS} reduces to 
\begin{align}
\label{eq:Ss}
S_{\mu\nu} = 2 C \kappa^2  \left(2 s_\mu s_\nu-s^2 g_{\mu \nu}\right)\,,\quad C=-\frac{1}{3}\left[\frac{8 A}{A^2+16 b_1^2}+\frac{B}{\Delta}\right]\,,
\end{align}
with
\begin{align}
    A &= 4+2 a_1+a_2, \qquad B = -8+2 a_1+a_2+3 a_3\,,\\
    \Delta &=2\left(a_1-a_2-1\right)\left(2 a_1+a_2+3 a_3-8\right)+\left(4 b_1-3 b_2\right)^2\,.
\end{align}

\subsection{Coarse-grained spin moments and spin-correlation anisotropy}

We define the coarse-grained (macroscopic) second moments
\begin{align}
\label{eq:defq}
Q_{\mu\nu}\equiv \langle s_\mu s_\nu\rangle,
\qquad
q\equiv\langle s_\alpha s^\alpha\rangle,
\end{align}
so that $u^\mu Q_{\mu\nu}=0$ and $q=h^{\mu\nu}Q_{\mu\nu}$ where $h_{\mu\nu}=g_{\mu\nu}+u_\mu u_\nu$. In the unpolarized (isotropic) case one assumes that the medium is isotropic on the coarse-graining scale, i.e. spins are randomly oriented with no preferred direction induced by magnetic fields, rotation, ferromagnetic ordering, or other microscopic effects. Rotational invariance in the local rest space then fixes the form of
\(Q_{\mu\nu}\) uniquely, since the only invariant spatial rank-two tensor is \(h_{\mu\nu}\)
\begin{align}
Q_{\mu\nu}=\frac{1}{3}\,q\,h_{\mu\nu}.
\end{align}
In the anisotropic case one introduces a unit spatial direction \(n^\mu\)
\begin{align}
u_\mu  n^\mu=0,
\qquad
 n_\mu  n^\mu=1\,.
\end{align}
If the state is axisymmetric around this direction ${n}^\mu$, then the only rank-2 spatial tensors invariant under rotations about ${n}$ are the spatial metric $h_{\mu \nu}$, and the dyad ${n}_\mu {n}_\nu$. The most general axisymmetric form is therefore 
\begin{align}
Q_{\mu \nu}=C_1 h_{\mu \nu}+C_2 {n}_\mu {n}_\nu,
\end{align}
with two scalars $C_1, C_2$. Using Eq.~\eqref{eq:defq} gives
\begin{align}
q=h^{\mu \nu}\left(C_1 h_{\mu \nu}+C_2 {n}_\mu {n}_\nu\right)=3 C_1+C_2\,.
\end{align}
Therefore only one free parameter remains once \(q\) is fixed. It is convenient to parametrize that remaining freedom by a dimensionless spin-correlation anisotropy parameter \(\chi\) by setting
\begin{align}
C_2=\chi^2 q, \quad C_1=\frac{1-\chi^2}{3} q,
\end{align}
which automatically satisfies $3 C_1+C_2=q$ and keeps $Q$ positive semidefinite in the local rest space for \(0\leq\chi\leq1\).
That choice yields
\begin{align}
Q_{\mu \nu}=\frac{1}{3}\left(1-\chi^2\right) q h_{\mu \nu}+\chi^2 q\, {n}_\mu {n}_\nu .
\end{align}
The parameter \(\chi\) should be interpreted with some care. Since it enters through the second moment \(Q_{\mu\nu}=\langle s_\mu s_\nu\rangle\), it does not necessarily describe a nonzero macroscopic spin density \(\langle s_\mu\rangle\).
Rather, it parametrizes an anisotropy of the local spin-spin correlations. In the static and spherically symmetric configurations considered below, the preferred direction \(n^\mu\) will be chosen as the radial direction. Thus, schematically in a local frame adapted to the radial direction, \(\chi\neq0\) corresponds to
\begin{align}
\langle s_r^2\rangle
\neq
\langle s_\theta^2\rangle
=
\langle s_\phi^2\rangle ,
\end{align}
not to a net radial spin polarization.

Since the configurations considered here are static and non-rotating, such a radial anisotropy is not attributed to rotational polarization. It should instead be viewed as a phenomenological description of possible microscopic ordering effects in dense matter, for example anisotropic spin correlations associated with crustal, pasta, or other ordered phases. In the absence of a microscopic mechanism producing such an ordering, the natural choice is the unpolarized limit \(\chi=0\).

\subsection{Effective fluid variables}

Having defined $Q_{\mu\nu}$ we can obtain the coarse-grained torsion-induced contribution \eqref{eq:Ss}
\begin{align}
\langle S_{\mu\nu} \rangle
=-2 C \kappa^2\Bigl(q\, g_{\mu\nu}-2 Q_{\mu\nu}\Bigr)=-2 C \kappa^2 q\left(\frac{1+2\chi^2}{3}h_{\mu\nu}-u_\mu u_\nu-2\chi^2  n_\mu n_\nu\right)
\end{align}
so we obtain different pressures parallel and orthogonal to the polarization
direction. In a spherically symmetric configuration where \(n^\mu\) is chosen as
the radial unit vector, these correspond to radial and tangential pressures. The coarse grained Einstein equation is therefore sourced by an effective fluid
\begin{align}
\overset{\,*}{R}_{\mu\nu} -\frac{1}{2} \overset{\,*}{R} g_{\mu\nu} = \kappa T_{\mu\nu}^{\text{(eff)}}\,,\qquad     T_{\mu\nu}^{\text{(eff)}} = T_{\mu\nu}+\frac{1}{\kappa} \langle S_{\mu\nu} \rangle\,.
\end{align}
Assuming that the ordinary matter contribution is described by a perfect fluid,
\begin{align}
T_{\mu\nu}=(\rho+p)u_\mu u_\nu+p g_{\mu\nu},
\end{align}
the effective density and pressures are
\begin{align}
\label{eq:matter}
\rho_{\mathrm{eff}}=\rho+2C \kappa q, \quad p_{r, \mathrm{eff}}=p-2 C \kappa q \frac{1-4\chi^2}{3}, \quad p_{t, \mathrm{eff}}=p-2 C \kappa q \frac{1+2\chi^2}{3}\,.
\end{align}

\subsection{Microscopic spin content and density scale}

In the spin-fluid approximation, the spatial spin-density vector is written as
\begin{align}
\mathbf{s}=n \mathbf{S}
\end{align}
where $n$ is the particle number density and $S^i$ is the intrinsic spin per particle, measured in the local rest frame of the fluid. We therefore
define
\begin{align}
q=\langle\mathbf{s}^2\rangle=\left\langle(n \mathbf{S})^2\right\rangle=n^2\langle\mathbf{S}^2\rangle .
\end{align}
Since the magnitude of the intrinsic spin is set by \(\hbar\), we parametrize the
averaged spin-density scalar as
\begin{align}
q=\alpha(\hbar c n)^2
\end{align}
where the factor of \(c\) has been restored. The dimensionless coefficient \(\alpha\) depends on the microscopic averaging prescription and on the normalization of the spin-density scalar. For example, the classical spin-vector normalization \(|\mathbf{S}|=\hbar/2\) gives \(\alpha=1/4\), whereas other conventions used in the literature, such as those of \cite{Nurgalev:1983vc,Poplawski:2023uhn}, lead to a different numerical factor.

Since only the product \(\alpha C\) enters the effective fluid variables, we choose the normalization \(\alpha=1/4\) in what follows.
\begin{align}
\label{eq:density}
& \rho_{\mathrm{eff}}=\rho+\frac{1}{2} C \kappa (\hbar c n)^2\,,\\
\label{eq:Pr}
& p_{r, \mathrm{eff}}=p-\frac{1}{2} C \kappa (\hbar c n)^2 \frac{1-4\chi^2}{3}\,,\\
\label{eq:Pt}
& p_{t, \mathrm{eff}}=p-\frac{1}{2} C \kappa (\hbar c n)^2 \frac{1+2\chi^2}{3}\,.
\end{align}
For \(C=\mathcal O(1)\), the torsion-induced correction is therefore
spin-squared and carries the usual \(\hbar^2\) suppression familiar from Einstein--Cartan theory \cite{Hehl:1976kj}. In that case the effect becomes relevant only at very large densities. In the present theory, however, the coefficient \(C\) is not fixed to its Einstein--Cartan value. It is a function of the dimensionless quadratic-torsion couplings \(a_i\) and \(b_i\), and is therefore not suppressed by a new mass scale. For this reason we shall also treat the effective spin-spin coupling phenomenologically. Equivalently, one may write
\begin{align}
C=\frac{N}{\hbar^2},
\end{align}
so that the explicit \(\hbar^2\) factor in \(q\) is absorbed into the
dimensionful parameter \(N\). This parametrization should not be understood as a generic prediction of the theory, but as a way of exploring regions of the dimensionless coupling space in which the effective spin-spin interaction is enhanced relative to Einstein--Cartan theory.

\subsection{Einstein--Cartan limit and spin equation of state}

As seen before, the Einstein--Cartan theory is obtained from the previous theory by setting
\begin{align}
a_1=a_2=a_3=b_1=b_2=0 .
\end{align}
For the Weyssenhoff fluid, the general result \eqref{eq:Ss} then gives
\begin{align}
A=4,
\qquad
B=-8,
\qquad
\Delta=16,
\qquad
C=-\frac{1}{2}.
\end{align}
Therefore the torsion-induced contribution becomes
\begin{align}
    S_{\mu\nu}=\kappa^2\left[-s_{\alpha\beta}s^{\alpha\beta}\left(u_\mu u_\nu+\frac{1}{2}g_{\mu\nu}\right)+2s_{\mu\alpha}s_\nu{}^\alpha\right]
\end{align}
or using the pseudovector \eqref{eq:decomp}
\begin{align}
\label{eq:SsEC}
S_{\mu\nu}
=
\kappa^2
\left(
s^2 g_{\mu\nu}
-
2s_\mu s_\nu
\right).
\end{align}
In the unpolarized case, \(\chi=0\), we obtain
\begin{align}
\rho_{\mathrm{eff}}
=
\rho-\kappa q,
\qquad
p_{\mathrm{eff}}
=
p+\frac{\kappa q}{3},
\end{align}
or equivalently
\begin{align}
\rho_{\mathrm{spin}}
=
-\kappa q,
\qquad
p_{\mathrm{spin}}
=
\frac{\kappa q}{3},
\qquad
w_{\mathrm{spin}}\equiv \frac{p_{\mathrm{spin}}}{\rho_{\mathrm{spin}}}
=
-\frac{1}{3}.
\end{align}
This result is consistent with \cite{Morawetz:2020lea}, but differs from other works, such as \cite{Obukhov:1987yu,Brechet:2007cj,deBerredo-Peixoto:2009apk,Boehmer:2006gd, Jockel:2024fps}, where the spin sector is commonly interpreted as an effective stiff component, $w_{\mathrm{spin}}=1$. This difference can be traced to the identification of the effective energy-momentum tensor after torsion has been eliminated. In the tetrad formulation, the Einstein--Cartan equations couple to the canonical energy-momentum tensor, which is generally not symmetric in the presence of spin. After eliminating torsion, the equations can instead be written in a purely Levi-Civita form by introducing a Belinfante--Rosenfeld improved tensor. Therefore, the effective density and pressure assigned to the spin sector depend on which part of the resulting equation is identified as the matter tensor and which part is isolated as an effective spin contribution. 

In the metric approach used here, this ambiguity does not arise, because the stress-energy tensor is defined directly by metric variation. The torsion-induced contribution then gives \(w_{\mathrm{spin}}=-1/3\) in the unpolarized case. More details are given in \hyperref[appendixD]{Appendix~\ref*{appendixD}}.

It is important to note that Ref.~\cite{Medina:2018rnl} discusses an ambiguity in the identification of the physical energy-momentum tensor, which in turn affects the interpretation of the torsion-induced contributions. In the present work, we follow the variational approach and identify the matter energy-momentum tensor with the tensor obtained by variation of the matter action with respect to the metric.


\section{Stellar structure equations}
\subsection{Effective anisotropic TOV equations}

To construct static and spherically symmetric stellar configurations, we consider the Schwarzschild-like line element
\begin{align}
ds^2
=
-e^{\nu}dt^2
+
e^{\lambda}dr^2
+
r^2 d\theta^2
+
r^2\sin^2\theta\, d\phi^2 .
\end{align}
The structure equations follow from the conservation of the effective energy--momentum tensor together with the Einstein equations for the metric above. Since the torsion-induced contribution behaves as an anisotropic fluid, the equilibrium equations take the form of modified TOV equations,
\begin{equation}
\begin{aligned}
\frac{dp_{r,\mathrm{eff}}}{dr}
&=
-\frac{
(\rho_{\mathrm{eff}}+p_{r,\mathrm{eff}})
\left(
m+4\pi r^3 p_{r,\mathrm{eff}}
\right)
}
{r(r-2m)}
-\frac{2}{r}
\left(
p_{r,\mathrm{eff}}-p_{t,\mathrm{eff}}
\right)
\\
&=
-\frac{
(\rho_{\mathrm{eff}}+p_{r,\mathrm{eff}})
\left(
m+4\pi r^3 p_{r,\mathrm{eff}}
\right)
}
{r(r-2m)}
-
\frac{8C\kappa q\chi^2}{r},
\end{aligned}
\end{equation}
together with
\begin{align}
\frac{dm}{dr}
=
4\pi r^2 \rho_{\mathrm{eff}},
\end{align}
and
\begin{align}
\frac{d\nu}{dr}
=
2
\frac{
m+4\pi r^3 p_{r,\mathrm{eff}}
}
{r(r-2m)}.
\end{align}
The boundary conditions are imposed on the effective thermodynamic variables.
\begin{align}
p_{r,\mathrm{eff}}(0)=p_c,
\qquad
\rho_{\mathrm{eff}}(0)=\rho_c,
\qquad
m(0)=0,
\end{align}
We also require vanishing anisotropy at the origin,
\begin{align}
\left(
p_{r,\mathrm{eff}}-p_{t,\mathrm{eff}}
\right)\Big|_{r=0}=0.
\end{align}
which imposes $\chi(r=0)=0$. The stellar surface is defined by the condition
\begin{align}
p_{r,\mathrm{eff}}(R)=0,
\end{align}
For weak polarization, \(\chi<1/2\), Eq. \eqref{eq:Pr} implies that the
spin contribution lowers the radial effective pressure only if \(C>0\). Indeed,
the surface condition \(p_{r,\mathrm{eff}}(R)=0\) gives
\begin{align}
p(R)
=
\frac12 C\kappa(\hbar c n(R))^2
\frac{1-4\chi^2}{3}.
\end{align}
Thus, for a finite surface density \(n(R)\neq0\), a non-negative matter pressure at the surface requires \(C>0\). If \(C<0\), the vanishing of \(p_{r,\mathrm{eff}}\) would require the matter pressure itself to become negative. In Einstein--Cartan theory, with our conventions, \(C=-1/2\). Therefore, within the present Weyssenhoff-fluid model and for weak polarization, Einstein--Cartan theory does not give a physically acceptable neutron-star configuration with finite surface density.

For \(C>0\), the spin contribution is attractive: it increases the effective energy density, lowers the effective radial pressure, and produces an inward anisotropic force. We therefore expect the resulting equilibrium configurations to be more compact than the corresponding perfect-fluid stars.

Once the stellar radius \(R\) has been determined from the condition \(p_{r,\mathrm{eff}}(R)=0\), the total gravitational mass is obtained by evaluating the mass function at the surface,
\begin{align}
M=m(R).
\end{align}

Strictly speaking, the matching of a spin fluid interior to a vacuum exterior in a torsion theory requires the corresponding junction conditions. In Einstein--Cartan theory these conditions were derived in \cite{Arkuszewski:1975fz}. They show that, for a Weyssenhoff fluid, the surface condition is modified by the spin sector and the ordinary matter pressure need not vanish at the boundary. In the effective Riemannian formulation used here, we impose the vanishing of the effective radial pressure,
\begin{align}
p_{r,\mathrm{eff}}(R)=0,
\end{align}
which is the condition required for matching the interior solution to a vacuum exterior without an additional surface layer. A full derivation of the junction conditions in the present quadratic-torsion theory is beyond the scope of this work.

\subsection{Evolution equations in matter variables}

For numerical integration it is convenient to rewrite the system in terms of the matter pressure \(p\) and energy density \(\rho\), rather than the effective variables. Substituting the effective quantities into the TOV equations yields
\begin{equation}
\begin{aligned}
\frac{dp}{dr}
={}&
-
\Bigg(
1
-
2C\kappa
\frac{1-4\chi^2}{3}
\frac{d\rho}{dp}
\frac{dq}{d\rho}
\Bigg)^{-1}
\Bigg[
\frac{
\left(
p+\rho
+
4C\kappa q
\frac{1+2\chi^2}{3}
\right)
\left(
m
+
4\pi r^3
\left(
p
-
2C\kappa q
\frac{1-4\chi^2}{3}
\right)
\right)
}
{r(r-2m)}
\\
&\qquad\qquad
+
\frac{8C\kappa q\chi^2}{r}
+
\frac{16C\kappa q\chi}{3}
\frac{d\chi}{dr}
\Bigg].
\end{aligned}
\end{equation}
The mass equation becomes
\begin{align}
\frac{dm}{dr}
=
4\pi r^2 \rho
+
8\pi r^2 C\kappa q,
\end{align}
while the metric function satisfies
\begin{align}
\frac{d\nu}{dr}
=
2
\frac{
m+4\pi r^3 p
}
{r(r-2m)}
-
2
\frac{
8\pi r^3 C\kappa q
}
{r(r-2m)}
\left(
\frac{1-4\chi^2}{3}
\right).
\end{align}

\subsection{Regularity and polarization profiles}

Regularity at the center of a spherically symmetric star requires the pressure anisotropy to vanish,
\begin{align}
p_{r,\text{eff}}-p_{t,\text{eff}} \rightarrow 0
\qquad \text{as} \qquad r\rightarrow 0.
\end{align}
In the present model, the anisotropy is given by
\begin{align}
p_{r,\text{eff}}-p_{t,\text{eff}} = 4C \kappa q \chi^2 .
\end{align}
Since $q=\frac{1}{4}(\hbar c n)^2$ is determined by the number density of matter, it generally remains finite and nonvanishing at the stellar center. Regularity therefore requires the polarization fraction to satisfy
\begin{align}
\chi \rightarrow 0
\qquad \text{as} \qquad r\rightarrow 0.
\end{align}
This condition is also physically well motivated. Near the stellar core, the temperature is expected to be sufficiently high that the microscopic spins become effectively randomly oriented, suppressing any macroscopic polarization.

We consider the following radial profiles for the polarization fraction: 
\begin{align}
\text{(i)} \quad & \chi =0, \\
\text{(ii)} \quad & \chi =\chi_0 \frac{r}{R}, \\
\text{(iii)} \quad & \chi =\chi_0 \tanh\!\left(\frac{r}{R}\right),
\end{align}
where $\chi_0<1/2$ to satisfy the weak polarization condition. The first case corresponds to an isotropic configuration, whereas the latter two describe smoothly increasing polarization away from the stellar center.

\section{Numerical results}
\subsection{Equation of state and parameter range}

To describe neutron-star matter, we employ the DD2 equation of state, including electrons, obtained from the CompOSE database \cite{Hempel_2010}.

We now specify the parameter range used in the numerical integration. As discussed above, for weak spin-correlation anisotropy, \(\chi<1/2\), physically acceptable stellar configurations with finite surface density require \(C>0\). In the isotropic limit, \(\chi=0\), Eq.~\eqref{eq:Pr} reduces to
\begin{align}
p_{r,\mathrm{eff}}
=
p
-
\frac{1}{6}C\kappa(\hbar c n)^2 .
\end{align}
Using the parametrization
\begin{align}
C=\frac{N}{\hbar^2},
\end{align}
this becomes
\begin{align}
p_{r,\mathrm{eff}}
=
p
-
\frac{1}{6}N\kappa c^2 n^2 .
\end{align}
The condition \(p_{r,\mathrm{eff}}>0\) therefore gives the approximate upper bound
\begin{align}
N
<
\frac{6p}{\kappa c^2 n^2}.
\end{align}

We estimate this bound using the range of pressures encountered in the numerical integration. In geometrized units, \(G=c=1\), the largest central pressure considered is of order
\begin{align}
p_c \sim 10^{-2}\ \mathrm{km}^{-2},
\end{align}
whereas the integration is stopped once the pressure falls below
\begin{align}
p \sim 10^{-15}\ \mathrm{km}^{-2}.
\end{align}
For the DD2 equation of state, these pressures correspond approximately to number densities
\begin{align}
n \sim 10^{54}\ \mathrm{km}^{-3},
\qquad
n \sim 10^{46}\ \mathrm{km}^{-3},
\end{align}
respectively. Substitution into the previous inequality gives the rough estimates
\begin{align}
N \lesssim 10^{-110}\ \mathrm{km}^4,
\qquad
N \lesssim 10^{-107}\ \mathrm{km}^4 .
\end{align}
We therefore use
\begin{align}
N \sim 10^{-114}\ \mathrm{km}^4,
\end{align}
which lies safely below the estimated bounds and leads to regular, physically acceptable numerical solutions.

In geometrized units, with length measured in km, the Planck constant is
\begin{align}
\hbar \simeq 2.61\times 10^{-76}\ \mathrm{km}^2,
\qquad
\hbar^2 \simeq 6.8\times 10^{-152}\ \mathrm{km}^4 .
\end{align}
Therefore the value \(N\sim 10^{-114}\ \mathrm{km}^4\) corresponds to the dimensionless coupling
\begin{align}
C=\frac{N}{\hbar^2}
\sim 10^{37}.
\end{align}
Thus the relevant hierarchy is not the smallness of the dimensionful parameter
\(N\), but the large positive value of the dimensionless coupling \(C\). The
numerical analysis below should therefore be understood as an exploration of the
positive-\(C\) branch of the quadratic-torsion parameter space.

It is useful to translate the large value of $C$ back into the original dimensionless couplings of the quadratic-torsion theory. From Eq.~\eqref{eq:Ss}, an enhanced value $C\gg 1$ is not obtained for generic order-one couplings. It corresponds instead to regions of parameter space where the algebraic torsion operator is close to being non-invertible. There are two possible ways this can occur. The trace--axial sector becomes nearly degenerate when
\begin{align}
\Delta
=
2 (a_1-a_2-1)(2a_1+a_2+3 a_3-8)
+
(4 b_1-3 b_2)^2
\simeq 0 .
\end{align}
The second possibility corresponds to a near-degeneracy of the purely tensor sector, namely
\begin{align}
(4+2a_{1}+a_{2})^{2}+16b_{1}^{2} \simeq 0 ,
\end{align}
which is equivalent to
\begin{align}
4+2a_{1}+a_{2}\simeq 0,
\qquad
b_{1} \simeq 0 .
\end{align}
The numerical models considered below should therefore be understood as probing an enhanced-response region of the algebraic torsion parameter space, while the exactly degenerate cases are excluded.

\subsection{Mass--radius relations}

\begin{figure}[h]
    \centering
    \includegraphics[width=0.48\textwidth]{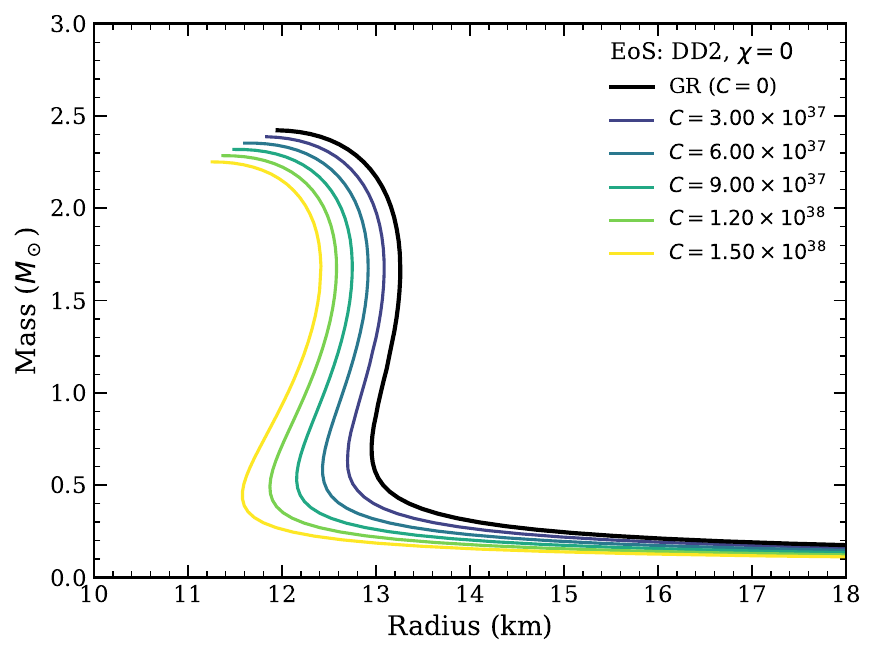}
\caption{Mass--radius relation for the DD2 equation of state in the isotropic
case, \(\chi=0\), for different values of \(C\). The GR sequence corresponds to
\(C=0\).}
    \label{fig:MRplot_DD2_zero}
\end{figure}

We first consider isotropic configurations, \(\chi=0\), and study the effect of the coupling parameter \(C\) on the mass--radius relation. The resulting sequences are shown in Fig.~\ref{fig:MRplot_DD2_zero}.

As \(C\) increases, the mass--radius curve is displaced toward smaller radii in the high-density part of the sequence. Thus, for comparable gravitational masses, the torsion-corrected configurations are more compact than their GR counterparts. At the same time, the maximum supported mass decreases as the torsion contribution becomes stronger, with the peak of the sequence shifting toward both lower masses and smaller radii. 

The different curves remain nearly indistinguishable on the low-mass branch, where the central pressure and central density are lower. This is consistent with the structure of the correction itself, since the spin contribution scales as \(q\propto n^2\). The torsion correction is therefore suppressed at low densities and becomes significant only toward the high-density part of the sequence, near the maximum-mass region.

\subsection{Binding energy and baryonic mass}

We next compare the gravitational mass \(M\) with the baryonic rest mass \(M_B\). The latter is defined from the conserved baryon number as
\begin{align}
M_B=m_b N_B, \qquad N_B = 4\pi \int_0^R n(r) \left(1-\frac{2m(r)}{r}\right)^{-1/2} r^2\,dr ,
\end{align}
where \(m_b\) is the baryon rest mass and \(n(r)\) is the baryon number density. The factor \((1-2m/r)^{-1/2}\) is the proper-volume correction for the spherically symmetric metric.

\begin{figure}[h]
    \centering
    \includegraphics[width=0.48\textwidth]{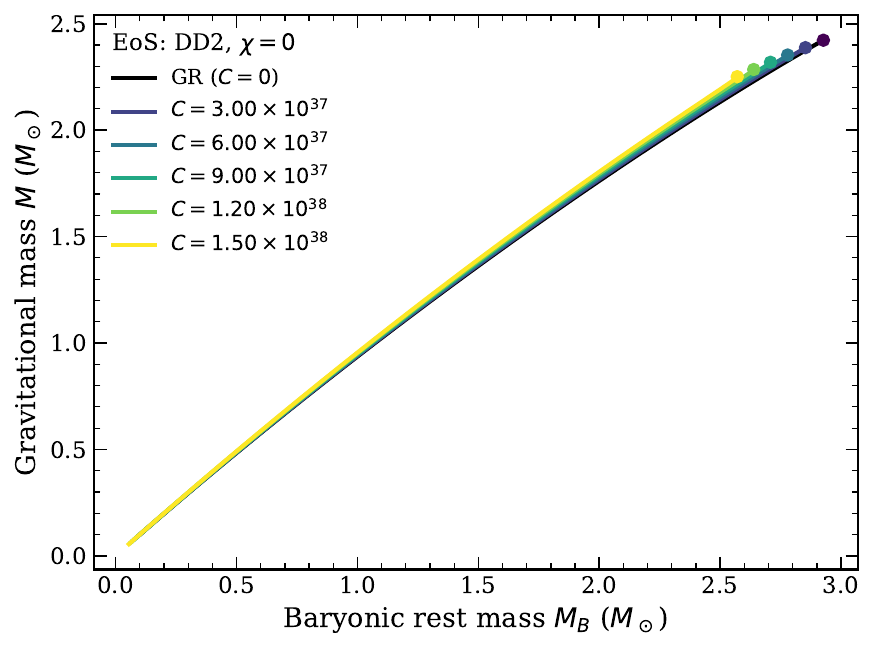}
    \caption{Gravitational mass \(M\) as a function of the baryonic rest mass
\(M_B\) for the DD2 equation of state in the isotropic case, \(\chi=0\), for
different values of \(C\). The GR sequence corresponds to \(C=0\).}
    \label{fig:M_vs_MB_isotropic}
\end{figure}

The corresponding curves are shown in Fig.~\ref{fig:M_vs_MB_isotropic}. For all values of \(C\), one finds \(M<M_B\) along the sequence, so the configurations are gravitationally bound. The difference
\begin{align}
E_b=M_B-M
\end{align}
is the binding energy, or equivalently the energy released in forming the star from the same baryonic rest mass.

For a fixed baryonic mass, the torsion-corrected configurations have a slightly larger gravitational mass than the GR configuration. Hence the binding energy is reduced relative to GR. The effect remains small over most of the sequence and becomes more visible close to the high-mass end, consistently with the density dependence of the torsion correction.

\subsection{Anisotropy and polarization profiles}

We have also integrated the modified TOV equations for the non-isotropic polarization profiles (ii) and (iii) defined in Section~4.3. The results, however, are largely indistinguishable from the isotropic case at the level of the mass--radius relation. For all values of $C$ considered and across both polarization profiles, the deviation in gravitational mass at fixed radius relative to the $\chi = 0$ sequence is of order $0.005$--$0.01\,M_\odot$. We therefore do not display separate mass--radius plots for the anisotropic cases, as they would be visually identical to Fig.~\ref{fig:MRplot_DD2_zero}.

For the smooth, centrally vanishing profiles considered here, $\chi$ remains well below unity throughout most of the stellar interior, and the net contribution to the mass and pressure profile away from isotropy is negligible.

Given the negligible effect of polarization on the mass--radius relation, we restrict the quantitative analysis to the isotropic case $\chi = 0$. The anisotropic profiles are retained in the formalism because they illustrate the general structure of the effective fluid, but they do not yield new, stable stellar sequences that are observationally distinguishable from the isotropic ones.

\section{Discussions and conclusions}
\label{sec:conclusion}

We have studied neutron-star equilibrium in an algebraic sector of Poincaré gauge gravity with quadratic torsion. The action contains the Einstein--Hilbert term supplemented by the independent parity-even and parity-odd quadratic torsion invariants. Since no kinetic term for torsion is included, the contorsion remains non-propagating and its field equation can be solved algebraically in terms of the spin current.

After eliminating the contorsion, the theory admits an effective Riemannian description. The metric field equations take the form of the Einstein equations sourced by the usual matter stress-energy tensor plus a spin-squared correction. For a Weyssenhoff fluid satisfying the Frenkel condition, this correction reduces to
\begin{align}
S_{\mu\nu}
=
2C\kappa^2
\left(
2s_\mu s_\nu-s^2g_{\mu\nu}
\right),
\end{align}
where the coefficient $C$ is a function of the dimensionless quadratic-torsion couplings. This is the main difference with respect to Einstein--Cartan theory: the microscopic spin density still carries the usual $\hbar^2$ suppression, but the coefficient multiplying the spin-squared term is no longer fixed to its Einstein--Cartan value.

We then introduced coarse-grained spin moments in order to describe both isotropic and anisotropic spin correlations. For an isotropic spin distribution, the torsion-induced correction behaves as a perfect-fluid contribution. For an axisymmetric spin-correlation tensor, it generates an effective pressure anisotropy. The resulting effective fluid variables are
\begin{align}
\rho_{\mathrm{eff}} &= \rho+2C\kappa q,\\
p_{r,\mathrm{eff}} &= p-2C\kappa q\,\frac{1-4\chi^2}{3},\\
p_{t,\mathrm{eff}} &= p-2C\kappa q\,\frac{1+2\chi^2}{3}.
\end{align}
These expressions show explicitly how the sign of $C$ determines whether the spin sector increases or decreases the effective pressure support.

The Einstein--Cartan limit was analyzed separately. Setting
\begin{align}
a_1=a_2=a_3=b_1=b_2=0
\end{align}
gives $C=-1/2$. With the metric definition of the stress-energy tensor used in this work, the unpolarized spin contribution is then
\begin{align}
\rho_{\mathrm{spin}}=-\kappa q,
\qquad
p_{\mathrm{spin}}=\frac{\kappa q}{3},
\qquad
w_{\mathrm{spin}}=-\frac{1}{3}.
\end{align}
This differs from the commonly quoted stiff contribution $w_{\mathrm{spin}}=1$. The difference is not a contradiction, but comes from the identification of the effective matter tensor after torsion has been eliminated. In the metric formulation adopted here, the matter stress-energy tensor is defined by variation with respect to the metric, and the remaining spin-squared contribution is isolated accordingly.

Using the effective anisotropic fluid, we derived the modified Tolman--Oppenheimer--Volkoff equations. The stellar surface was defined by the vanishing of the effective radial pressure, $p_{r,\mathrm{eff}}(R)=0$. This choice is natural in the effective Riemannian formulation. Nevertheless, the matching of a spin-fluid interior to a vacuum exterior in a torsion theory is subtle, and a complete derivation of the junction conditions for the present quadratic-torsion theory remains an open extension of the present analysis.

We solved the stellar structure equations numerically using the DD2 equation of state. For the positive effective spin-spin coupling branch considered here, the torsion correction increases the compactness of the configurations. The high-density part of the mass--radius sequence is shifted toward smaller radii, while the maximum supported mass decreases. The low-mass branch is almost unchanged, as expected from the scaling $q\propto n^2$, which suppresses the correction at low densities.

We also computed the relation between gravitational mass and baryonic rest mass. Along the sequences studied here, the configurations remain gravitationally bound, $M<M_B$. However, at fixed baryonic mass, the torsion-corrected configurations have a slightly larger gravitational mass than the corresponding general-relativistic configurations. The binding energy is therefore reduced, with the effect becoming visible mainly near the high-mass end of the sequence.

Finally, we considered simple weak-polarization profiles satisfying regularity at the center. For these smooth profiles, the effect of spin-correlation anisotropy on the mass--radius relation is very small, with deviations in the gravitational mass at fixed radius of order $0.005$--$0.01\,M_\odot$. Thus, within the class of profiles considered here, the dominant observable effect comes from the isotropic spin-squared correction rather than from anisotropy.

The enhanced values of $C$ explored numerically should be interpreted phenomenologically. They correspond to regions of the quadratic-torsion parameter space where the algebraic torsion operator is close to a non-invertible limit, while the exactly degenerate cases are excluded. The results therefore show that algebraic quadratic torsion can produce measurable changes in neutron-star structure if the effective spin-spin response is sufficiently enhanced relative to the Einstein--Cartan value.

Several directions remain open for future work. A natural extension is to include kinetic terms for torsion, allowing for propagating degrees of freedom in the dynamical sector of Poincaré gauge gravity. It would also be worthwhile to study slowly rotating neutron stars and determine the impact of torsion on rotational properties such as the moment of inertia. Furthermore, extending the perturbation analysis to compute tidal Love numbers would provide access to observables directly relevant for gravitational-wave astronomy. Together with mass--radius measurements, such probes could offer new avenues for testing the role of torsion in the strong-field regime and constraining deviations from general relativity.

\section*{Acknowledgments}
The work of RG has been supported by ANID FONDECYT Regular No. 1262002 (Chile).

\appendix

\renewcommand*{\sectionformat}{\appendixname~\thesection\autodot\enskip}


\section{Nieh--Yan identity and Holst term}
\label{appendixA}

From the second Bianchi identity, we have
\begin{align}
R_{[\mu \nu \rho]\sigma}=2\, \nabla_{[\mu} T_{\nu \rho]\sigma}-4\, T_{[\mu \nu}{ }^{\lambda} T_{\rho] \lambda \sigma}
\end{align}
we obtain
\begin{align}
\varepsilon^{\mu\nu\rho\sigma}\,R_{\mu\nu\rho\sigma} &= 2\, \varepsilon^{\mu\nu\rho\sigma}\,\nabla_{\mu} T_{\nu \rho \sigma}-4\, \varepsilon^{\mu\nu\rho\sigma}\,T_{\mu \nu}{ }^{\lambda} T_{\rho \lambda \sigma}
= 2\, \nabla_{\mu} \Bigl(\varepsilon^{\mu\nu\rho\sigma}\, T_{\nu \rho \sigma}\Bigr)-4\, \varepsilon^{\mu\nu\rho\sigma}\,T_{\mu \nu}{ }^{\lambda} T_{\rho \lambda \sigma}\\
&= \frac{2}{\sqrt{-g}} \partial_{\mu} \Bigl(\sqrt{-g}\,\varepsilon^{\mu\nu\rho\sigma}\, T_{\nu \rho \sigma}\Bigr)-4\, \varepsilon^{\mu\nu\rho\sigma}\,T_{\mu \nu}{ }^{\lambda} T_{\rho \lambda \sigma}
\end{align}
which gives the Nieh-Yan identity 
\begin{align}
\partial_{\mu} \Bigl(\sqrt{-g}\,\varepsilon^{\mu\nu\rho\sigma}\, T_{\nu \rho \sigma}\Bigr)=\sqrt{-g}\,\varepsilon^{\mu\nu\rho\sigma}\Bigl(2\, T_{\mu \nu}{ }^{\lambda} T_{\rho \lambda \sigma}+\frac{1}{2}R_{\mu\nu\rho\sigma}\Bigr)
\end{align}
Therefore, the Holst density becomes
\begin{align}
\varepsilon^{\mu\nu\rho\sigma}\,R_{\mu\nu\rho\sigma} &= 2\, \varepsilon^{\mu\nu\rho\sigma}\,\nabla_{\mu} T_{\nu \rho \sigma}-4\, \varepsilon^{\mu\nu\rho\sigma}\,T_{\mu \nu}{ }^{\lambda} T_{\rho \lambda \sigma} \\
&= 2\, \varepsilon^{\mu\nu\rho\sigma}\,\nabla_{\mu} T_{\nu \rho \sigma}-4\,\Bigl(\frac{1}{2} \varepsilon^{\mu \nu \rho \sigma} T_{\mu \nu}{ }^\lambda T_{\rho \sigma \lambda}+\varepsilon^{\mu \nu \rho \sigma} T_{\mu \nu \rho} T_\sigma\Bigr)
\end{align}
where we have used 
\begin{align}
    \varepsilon^{\mu\nu\rho\sigma}\,T_{\mu \nu}{ }^{\lambda} T_{\rho \lambda \sigma} = \frac{1}{2} \varepsilon^{\mu \nu \rho \sigma} T_{\mu \nu}{ }^\lambda T_{\rho \sigma \lambda} + \varepsilon^{\mu \nu \rho \sigma} T_{\mu \nu \rho} T_\sigma 
\end{align}
Indeed, starting from
\begin{align}
\varepsilon^{\mu \nu \rho \sigma} T_{\mu \nu}{ }^\lambda T_{\rho \lambda \sigma}=\varepsilon^{\mu \nu \rho \sigma} T_{\mu \nu}{ }^\lambda T_{\lambda \sigma \rho}\,,
\end{align}
one finds
\begin{align}
2\varepsilon^{\mu \nu \rho \sigma} T_{\mu \nu}{ }^\lambda T_{\rho \lambda \sigma}-\varepsilon^{\mu \nu \rho \sigma} T_{\mu \nu}{ }^\lambda T_{\rho \sigma \lambda}=\varepsilon^{\mu \nu \rho \sigma} T_{\mu \nu}{ }^\lambda C_{\rho \lambda \sigma}\,,~~~\text{with}~~ C_{\rho \lambda \sigma}\equiv T_{\rho \lambda \sigma}+T_{\lambda \sigma \rho}+T_{\sigma \rho \lambda}\,.
\end{align}
But the tensor $C_{\rho \lambda \sigma}$ is totally antisymmetric, it can then be written as
\begin{align}
    C_{\rho \lambda \sigma}=-\varepsilon_{\rho \lambda \sigma \alpha} U^\alpha\,,~~~\text{with}~~U^\alpha\equiv \frac{1}{6} \varepsilon^{\alpha \rho \lambda \sigma} C_{\rho \lambda \sigma}\,,
\end{align}
which implies
\begin{align}
2\varepsilon^{\mu \nu \rho \sigma} T_{\mu \nu}{ }^\lambda T_{\rho \lambda \sigma}-\varepsilon^{\mu \nu \rho \sigma} T_{\mu \nu}{ }^\lambda T_{\rho \sigma \lambda}=-\varepsilon^{\mu \nu \rho \sigma}  \varepsilon_{\rho \lambda \sigma \alpha} T_{\mu \nu}{ }^\lambda U^\alpha = 2\left(\delta_\lambda^\mu \delta_\alpha^\nu-\delta_\alpha^\mu \delta_\lambda^\nu\right) T_{\mu \nu}{ }^\lambda U^\alpha=-4 T_\alpha U^\alpha\,.
\end{align}
Furthermore, 
\begin{align}
U^\alpha=\frac{1}{6} \varepsilon^{\alpha \rho \lambda \sigma}\left(T_{\rho \lambda \sigma}+T_{\lambda \sigma \rho}+T_{\sigma \rho \lambda}\right)=\frac{1}{2} \varepsilon^{\alpha \mu \nu \rho} T_{\mu \nu \rho}\,.
\end{align}
This leads to
\begin{align}
2\varepsilon^{\mu \nu \rho \sigma} T_{\mu \nu}{ }^\lambda T_{\rho \lambda \sigma}-\varepsilon^{\mu \nu \rho \sigma} T_{\mu \nu}{ }^\lambda T_{\rho \sigma \lambda}=-2\varepsilon^{\alpha \mu \nu \rho} T_{\mu \nu \rho} T_\alpha\,,
\end{align}
and finally
\begin{align}
\varepsilon^{\mu \nu \rho \sigma} T_{\mu \nu}{ }^\lambda T_{\rho \lambda \sigma}=\frac{1}{2}\varepsilon^{\mu \nu \rho \sigma} T_{\mu \nu}{ }^\lambda T_{\rho \sigma \lambda}-\varepsilon^{\alpha \mu \nu \rho} T_{\mu \nu \rho} T_\alpha = \frac{1}{2}\varepsilon^{\mu \nu \rho \sigma} T_{\mu \nu}{ }^\lambda T_{\rho \sigma \lambda}+\varepsilon^{\mu \nu \rho \alpha} T_{\mu \nu \rho} T_\alpha\,.
\end{align}

\section{Spin-induced contribution}

For completeness, here we give the explicit forms of the tensors
$\mathcal{M}_{\mu\nu}$ and $\mathcal{T}$ that determine the torsion-induced
correction \(S_{\mu\nu}\) in Eq.~\eqref{eq:fullS}.

\label{appendixB}

\begin{align}
\mathcal{M}_{\mu\nu}
={}&
A_{1}\,
\epsilon_{\alpha\beta\gamma(\nu}
\tau_{\rho}{}^{\beta\gamma}
\tau^{\rho}{}_{\mu)}{}^{\alpha}
+
A_{2}\,
\tau_{\rho(\nu|\alpha|}
\tau^{\rho}{}_{\mu)}{}^{\alpha}
+A_{3}\,
\epsilon_{\alpha\beta\gamma(\nu}
\tau^{\alpha}{}_{\mu)}{}^{\rho}
\tau_{\rho}{}^{\beta\gamma}
+
A_{4}\,
\tau^{\rho}{}_{(\mu}{}^{\alpha}
\tau_{\alpha\nu)\rho}
\nonumber\\[1mm]
&+
A_{5}\,
\epsilon_{\alpha\beta\gamma(\nu}
\tau^{\alpha}{}_{\mu)}{}^{\rho}
\tau^{\beta}{}_{\rho}{}^{\gamma}
+
A_{6}\,
\epsilon_{\alpha\beta\gamma\delta}
\tau^{\alpha}{}_{(\mu}{}^{\beta}
\tau^{\gamma}{}_{\nu)}{}^{\delta}
+
A_{7}\,
\tau_{\mu\alpha\beta}\tau_{\nu}{}^{\alpha\beta}
+
A_{8}\,
\epsilon_{\alpha\beta\gamma\delta}
\tau^{\gamma}{}_{(\nu}{}^{\delta}
\tau_{\mu)}{}^{\alpha\beta}
\nonumber\\[1mm]
&+
A_{9}\,
\epsilon_{\alpha\beta\gamma(\nu}
\tau^{\beta}{}_{\mu)}{}^{\gamma}\tau^{\alpha}
+
A_{10}\,
\epsilon_{\alpha\beta\gamma(\nu}
\tau_{\mu)}{}^{\beta\gamma}\tau^{\alpha}
+
A_{11}\,
\tau_{\mu\nu\alpha}\tau^{\alpha}
+
A_{12}\,
\tau_{\mu}\tau_{\nu}
\nonumber\\[1mm]
&+
A_{13}\,
\epsilon_{\alpha\beta\gamma(\nu}
\tau^{\beta}{}_{\mu)}{}^{\gamma}\tilde{\tau}^{\alpha}
+
A_{14}\,
\epsilon_{\alpha\beta\gamma(\nu}
\tau_{\mu)}{}^{\beta\gamma}\tilde{\tau}^{\alpha}
+
A_{15}\,
\tau_{\mu\nu\alpha}\tilde{\tau}^{\alpha}
+
A_{16}\,
\tau_{(\mu}\tilde{\tau}_{\nu)}
+
A_{17}\,
\tilde{\tau}_{\mu}\tilde{\tau}_{\nu}
\nonumber\\[1mm]
&+A_{18} \, \epsilon_{\alpha\beta\gamma\delta}\,\tau_{\mu}{}^{\alpha\beta}\,\tau_{\nu}{}^{\gamma\delta}\,
+
A_{19}\,
\epsilon_{(\mu|\alpha\beta\gamma|}\,\tau_{\delta}{}^{\beta\gamma}\,\tau_{\nu)}{}^{\delta\alpha}
+
A_{20}\,
\tau_{\delta(\mu|\alpha|}\,\tau_{\nu)}{}^{\delta\alpha}
+
A_{21}\,
\epsilon_{(\mu|\alpha\beta\gamma|}\,\tau^{\beta}{}_{\delta}{}^{\gamma}\,\tau_{\nu)}{}^{\delta\alpha}
\end{align}

\begin{equation}
\begin{aligned}
A_{1}
&=
256(2a_{1}-a_{2})(4+2a_{1}+a_{2})\,b_{1}\Delta,
\qquad
A_{2}
=
-64(4+2a_{1}+a_{2})
\Big[
4a_{1}(2+a_{1})
-a_{2}(4+a_{2})
+8b_{1}^{2}
\Big]\Delta,
\\[2mm]
A_{3}
&=
128b_{1}
\Big[
-4a_{1}^{2}
+4a_{1}a_{2}
+a_{2}(16+3a_{2})
+8(2+b_{1}^{2})
\Big]\Delta,
\qquad
A_{4}
=
-32(4+2a_{1}+a_{2})^{2}
(4-2a_{1}+3a_{2})\Delta,
\\[2mm]
A_{5}
&=
-256b_{1}
\Big[
(4+2a_{1}+a_{2})^{2}
-8b_{1}^{2}
\Big]\Delta,
\qquad
A_{6}
=
-128(4+2a_{1}+a_{2})^{2}b_{1}\Delta,
\\[2mm]
A_{7}
&=
-32(4+2a_{1}+a_{2})
\Big[
4a_{1}(2+a_{1})
-a_{2}(4+a_{2})
-8b_{1}^{2}
\Big]\Delta,
\\[2mm]
A_{8}
&=
-64(-4+2a_{1}-3a_{2})
(4+2a_{1}+a_{2})b_{1}\Delta,
\qquad
A_{9}
=
256b_{1}\,\mathcal{P}_{1},
\qquad
A_{10}
=
-128b_{1}\,\mathcal{P}_{2},
\\[2mm]
A_{11}
&=
64\Big[(4+2a_{1}+a_{2})^{2}+16b_{1}^{2}\Big]\mathcal{A},
\qquad
A_{12}
=
32\Big[(4+2a_{1}+a_{2})^{2}+16b_{1}^{2}\Big]\mathcal{A},
\\[2mm]
A_{13}
&=
-32\,\mathcal{B},
\qquad
A_{14}
=
16\,\mathcal{C},
\qquad
A_{15}
=
16\,\mathcal{D},
\qquad
A_{16}
=
256b_{1}\,\mathcal{F},
\qquad
A_{17}
=
-2\,\mathcal{G},
\\[2mm]
A_{18}
&=
\frac{1}{2}A_{1},
\qquad
A_{19}
=
A_{3},
\\[2mm]
A_{20}
&=
64(4+2a_{1}+a_{2})
\Big[
(2a_{1}-3a_{2})(2a_{1}+a_{2})
-16(1+a_{2}+b_{1}^{2})
\Big]\Delta,
\\[2mm]
A_{21}
&=
256b_{1}
\Big(
-4-2a_{1}-a_{2}+4b_{1}
\Big)
\Big(
4+2a_{1}+a_{2}+4b_{1}
\Big)\Delta.
\end{aligned}
\end{equation}

\begin{align}
\mathcal{A}
={}&
4 a_{1}^{2}(-4+a_{3})
-2(1+a_{2})(4+a_{2})(-4+a_{3})
+16(2+a_{2})b_{1}^{2}
-8(4+a_{2})b_{1}b_{2}
+3(4+a_{2})b_{2}^{2}
\nonumber\\
&+
2a_{1}
\Big[
(2-a_{2})(-4+a_{3})
-8b_{1}b_{2}
+3b_{2}^{2}
\Big].
\end{align}

\begin{align}
\mathcal{P}_{1}
={}&
32 a_{1}^{4}
-8 a_{1}^{3}(16+2a_{2}-7a_{3})
+2(1+a_{2})(4+a_{2})
\Big[
2(-4+a_{2})(-2+a_{2})
+(-4+5a_{2})a_{3}
\Big]
\nonumber\\
&+
32(-4-7a_{2}+2(1+a_{2})a_{3})b_{1}^{2}
-256b_{1}^{4}
+4b_{1}
\Big[
-16+a_{2}(40+11a_{2})+80b_{1}^{2}
\Big]b_{2}
\nonumber\\
&-
3\Big[
-16+a_{2}(16+5a_{2})+32b_{1}^{2}
\Big]b_{2}^{2}+
4a_{1}^{2}
\Big[
-72+18a_{3}
-6a_{2}(-2+a_{2}+2a_{3})
+16b_{1}^{2}
-52b_{1}b_{2}
+21b_{2}^{2}
\Big]
\nonumber\\
&+
2a_{1}
\Big[
2a_{2}^{3}
+a_{2}^{2}(24-9a_{3})
-16a_{3}(3+2b_{1}^{2})
+32(4+7b_{1}^{2}-7b_{1}b_{2}+3b_{2}^{2})
\nonumber \\
&+a_{2}(240-84a_{3}+16b_{1}^{2}-8b_{1}b_{2}+6b_{2}^{2})
\Big].
\end{align}

\begin{align}
\mathcal{P}_{2}
={}&
32 a_{1}^{4}
-8 a_{1}^{3}(16+2a_{2}-7a_{3})
+
2(1+a_{2})(4+a_{2})
\Big[
2(-4+a_{2})(-2+a_{2})
+(-4+5a_{2})a_{3}
\Big]
\nonumber\\
&+
16(-16+a_{2}^{2}+7a_{2}(-3+a_{3})+7a_{3})b_{1}^{2}
-384b_{1}^{4}
+
4b_{1}
\Big[
-16+40a_{2}+11a_{2}^{2}+128b_{1}^{2}
\Big]b_{2}
\nonumber\\
&-
3\Big[
-16+16a_{2}+5a_{2}^{2}+56b_{1}^{2}
\Big]b_{2}^{2}
-
4a_{1}^{2}
\Big[
72-18a_{3}
+6a_{2}(-2+a_{2}+2a_{3})
-8b_{1}^{2}
+52b_{1}b_{2}
-21b_{2}^{2}
\Big]
\nonumber\\
&+
2a_{1}
\Big[
2a_{2}^{3}
+a_{2}^{2}(24-9a_{3})
-8a_{3}(6+7b_{1}^{2})
+16(8+19b_{1}^{2}-14b_{1}b_{2}+6b_{2}^{2})
\nonumber\\
&\qquad
+a_{2}(240-84a_{3}+24b_{1}^{2}-8b_{1}b_{2}+6b_{2}^{2})
\Big].
\end{align}


\begin{align}
\mathcal{B}
={}&
32 a_{1}^{5}
-16 a_{1}^{4}(6+2a_{2}-3a_{3})
+
2(1+a_{2})(4+a_{2})^{2}(4+3a_{2})(-8+a_{2}+3a_{3})
\nonumber\\
&+
8(4+a_{2})
\Big[
28(-4+a_{3})
+a_{2}(-92+2a_{2}+25a_{3})
\Big]b_{1}^{2}
-
128(20+4a_{2}-a_{3})b_{1}^{4}
\nonumber\\
&+
2(4+a_{2})b_{1}
\Big[
208+a_{2}(200+37a_{2})+400b_{1}^{2}
\Big]b_{2}
-
9(4+a_{2})
\Big[
16+a_{2}(16+3a_{2})+32b_{1}^{2}
\Big]b_{2}^{2}
\nonumber\\
&-
8a_{1}^{3}
\Big[
40-6a_{3}
+a_{2}(4+6a_{2}+9a_{3})
+16b_{1}^{2}
+22b_{1}b_{2}
-9b_{2}^{2}
\Big]
\nonumber\\
&+
4a_{1}^{2}
\Big[
4a_{2}^{3}
+a_{2}^{2}(24-9a_{3})
+32(5+7b_{1}^{2})
-8a_{3}(9+11b_{1}^{2})
\nonumber\\
&\qquad
+36b_{2}(-2b_{1}+b_{2})
+a_{2}(216-90a_{3}-16b_{1}^{2}+30b_{1}b_{2}-9b_{2}^{2})
\Big]
\nonumber\\
&+
2a_{1}
\Big[
13a_{2}^{4}
+3a_{2}^{3}(12+7a_{3})
+a_{2}^{2}(-24+90a_{3}+16b_{1}^{2}+126b_{1}b_{2}-45b_{2}^{2})
\nonumber\\
&\qquad
+8a_{2}
\Big[
32+14(-4+a_{3})b_{1}^{2}
+78b_{1}b_{2}
-27b_{2}^{2}
\Big]
-16
\Big[
6(-4+a_{3})
+8(-2+a_{3})b_{1}^{2}
+32b_{1}^{4}
\nonumber\\
&\qquad\qquad
-10b_{1}(3+5b_{1}^{2})b_{2}
+9(1+2b_{1}^{2})b_{2}^{2}
\Big]
\Big],
\end{align}


\begin{align}
\mathcal{C}
={}&
32 a_{1}^{5}
-16 a_{1}^{4}(6+2a_{2}-3a_{3})
+
2(1+a_{2})(4+a_{2})^{2}(4+3a_{2})(-8+a_{2}+3a_{3})
\nonumber\\
&+
8(4+a_{2})
\Big[
-144+6(-20+a_{2})a_{2}+(40+37a_{2})a_{3}
\Big]b_{1}^{2}
-
128(28+6a_{2}-a_{3})b_{1}^{4}
\nonumber\\
&+
2(4+a_{2})b_{1}
\Big[
208+a_{2}(200+37a_{2})+592b_{1}^{2}
\Big]b_{2}
-
9(4+a_{2})
\Big[
16+a_{2}(16+3a_{2})+48b_{1}^{2}
\Big]b_{2}^{2}
\nonumber\\
&-
8a_{1}^{3}
\Big[
40-6a_{3}
+a_{2}(4+6a_{2}+9a_{3})
+32b_{1}^{2}
+22b_{1}b_{2}
-9b_{2}^{2}
\Big]
+
2a_{1}
\Big[
13a_{2}^{4}
+3a_{2}^{3}(12+7a_{3})
\nonumber\\
&\qquad
+a_{2}^{2}(-24+90a_{3}+64b_{1}^{2}+126b_{1}b_{2}-45b_{2}^{2})
+8a_{2}
\Big[
32+4(-14+5a_{3})b_{1}^{2}
+78b_{1}b_{2}
-27b_{2}^{2}
\Big]
\nonumber\\
&\qquad
-16
\Big[
2a_{3}(3+7b_{1}^{2})
+8(-3-5b_{1}^{2}+6b_{1}^{4})
-2b_{1}(15+37b_{1}^{2})b_{2}
+9(1+3b_{1}^{2})b_{2}^{2}
\Big]
\Big]
\nonumber\\
&+
4a_{1}^{2}
\Big[
4a_{2}^{3}
+a_{2}^{2}(24-9a_{3})
+a_{2}(216-90a_{3}-16b_{1}^{2}+30b_{1}b_{2}-9b_{2}^{2})
\nonumber\\
&\qquad
+4
\Big[
40+80b_{1}^{2}
-2a_{3}(9+17b_{1}^{2})
+9b_{2}(-2b_{1}+b_{2})
\Big]
\Big],
\end{align}

\begin{align}
\mathcal{D}
={}&
1024b_{1}^{5}
-(4+2a_{1}+a_{2})^{4}b_{2}
-288(4+2a_{1}+a_{2})^{2}b_{1}^{2}b_{2}
-
1280b_{1}^{4}b_{2}
\nonumber\\
&+
128b_{1}^{3}
\Big[
32+8a_{1}^{2}
+a_{2}(18+a_{2}-2a_{3})
-2a_{3}
+2a_{1}(10+3a_{2}+a_{3})
+3b_{2}^{2}
\Big]
\nonumber\\
&+
4(4+2a_{1}+a_{2})^{2}b_{1}
\Big[
12a_{1}^{2}
+a_{2}(40-5a_{2}-16a_{3})
-4a_{1}(12+a_{2}-4a_{3})
+8(6-2a_{3}+3b_{2}^{2})
\Big],
\end{align}

\begin{align}
\mathcal{F}
={}&
192
+8a_{1}^{4}
-a_{2}^{4}
+a_{2}^{3}(3-4a_{3})
+4a_{1}^{3}(-6+a_{2}+4a_{3})
+
8a_{3}(-8+b_{1}^{2})
+2a_{2}^{2}(42-18a_{3}+8b_{1}^{2}-7b_{1}b_{2}+3b_{2}^{2})
\nonumber\\
&+
8a_{2}
\Big[
34+9b_{1}^{2}
+a_{3}(-12+b_{1}^{2})
-14b_{1}b_{2}
+6b_{2}^{2}
\Big]
-
4(4b_{1}-3b_{2})
\Big[
8b_{2}+b_{1}(-8+4b_{1}^{2}-b_{1}b_{2})
\Big]
\nonumber\\
&-
2a_{1}^{2}
\Big[
3a_{2}(2+a_{2})
-4(6a_{3}+2b_{1}^{2}-7b_{1}b_{2}+3(-6+b_{2}^{2}))
\Big]
-
a_{1}
\Big[
5a_{2}^{3}
+6a_{2}^{2}(-1+2a_{3})
\nonumber\\
&\qquad
+8(4+(-18+a_{3})b_{1}^{2}+28b_{1}b_{2}-12b_{2}^{2})
+8a_{2}(6a_{3}-5b_{1}^{2}+7b_{1}b_{2}-3(4+b_{2}^{2}))
\Big],
\end{align}


\begin{align}
\mathcal{G}
={}&
96a_{1}^{5}
+27a_{2}^{5}
-48a_{1}^{4}(8+3a_{2}-3a_{3})
+9a_{2}^{4}(8+9a_{3})
+
256
\Big[
-8(1+b_{1}^{2})(3+11b_{1}^{2})
+a_{3}(9+32b_{1}^{2}+5b_{1}^{4})
\nonumber\\
&
+36(b_{1}+3b_{1}^{3})b_{2}
-3(4+13b_{1}^{2})b_{2}^{2}
\Big]
-
64a_{2}
\Big[
180+448b_{1}^{2}+68b_{1}^{4}
-2a_{3}(36+71b_{1}^{2})
\nonumber\\
&
-12b_{1}(13+9b_{1}^{2})b_{2}
+3(18+13b_{1}^{2})b_{2}^{2}
\Big]
-16a_{1}^{3}
\Big[
9a_{2}(a_{2}+2a_{3})
+4(12+28b_{1}^{2}+6b_{1}b_{2}-3b_{2}^{2})
\Big]
\nonumber\\
&+
32a_{2}^{2}
\Big[
-124b_{1}^{2}
+11a_{3}(9+5b_{1}^{2})
+102b_{1}b_{2}
-36(6+b_{2}^{2})
\Big]
+8a_{1}^{2}
\Big[
15a_{2}^{3}
-9a_{2}^{2}(-8+a_{3})
\nonumber\\
&
-16(-24+9a_{3}-44b_{1}^{2}+23a_{3}b_{1}^{2}-6b_{1}b_{2})
-12a_{2}(4(-9+3a_{3}+b_{1}^{2})-10b_{1}b_{2}+3b_{2}^{2})
\Big]
\nonumber\\
&+
8a_{2}^{3}
\Big[
108a_{3}
+44b_{1}^{2}
+42b_{1}b_{2}
-3(52+5b_{2}^{2})
\Big]
+2a_{1}
\Big[
63a_{2}^{4}
+12a_{2}^{3}(16+9a_{3})
\nonumber\\
&
+64a_{2}
\Big[
a_{3}(9+16b_{1}^{2})
-2(4b_{1}-3b_{2})(5b_{1}-3b_{2})
\Big]
+24a_{2}^{2}
(-12+24a_{3}+20b_{1}^{2}+26b_{1}b_{2}-9b_{2}^{2})
\nonumber\\
&
+64
\Big[
12+(32-14a_{3})b_{1}^{2}
-68b_{1}^{4}
+12b_{1}(5+9b_{1}^{2})b_{2}
-3(6+13b_{1}^{2})b_{2}^{2}
\Big]
\Big],
\end{align}


\begin{align}
\mathcal{T}
={}&
16 (4 + 2 a_{1} + a_{2})
\Big[
4 a_{1} (2 + a_{1}) - a_{2} (4 + a_{2}) + 8 b_{1}^{2}
\Big] \Delta\,
\tau_{abc}\tau^{abc} -
64b_{1}\,\mathcal{H}\,
\tau_{a}\tilde{\tau}^{a}
+
\mathcal{I}\,
\tilde{\tau}_{a}\tilde{\tau}^{a}
\nonumber\\
&-64 (2 a_{1} - a_{2}) (4 + 2 a_{1} + a_{2}) b_{1} \Delta\,
\epsilon_{bcdf}\tau_{a}{}^{df}\tau^{abc} -16\Big[(4+2a_{1}+a_{2})^{2}+16b_{1}^{2}\Big]
\mathcal{A}\,
\tau_{a}\tau^{a}
\nonumber\\
&+
16 (4 + 2 a_{1} + a_{2})^{2}(4 - 2 a_{1} + 3 a_{2})\Delta\,
\tau_{abc}\tau^{bac} +
128(4+2a_{1}+a_{2})^{2}b_{1}\Delta\,
\epsilon_{acdf}\tau^{abc}\tau^{d}{}_{b}{}^{f}
\nonumber\\
&-
64 b_{1}
\Big[
(2 a_{1} + a_{2})(-2 a_{1} + 3 a_{2})
+16(1+a_{2}+b_{1}^{2})
\Big]\Delta\,
\epsilon_{acdf}\tau^{abc}\tau_{b}{}^{df}
\end{align}


\begin{align}
\mathcal{H}
={}&
48 a_{1}^{4}
-8 a_{1}^{3}(22+a_{2}-11a_{3})
+2(1+a_{2})(4+a_{2})(64-20a_{3}+a_{2}(-4+a_{2}+a_{3}))
\nonumber\\
&+
16\Big[
2(8+a_{3})+a_{2}(2+a_{2}+2a_{3})
\Big]b_{1}^{2}
-256b_{1}^{4}
+16b_{1}
\Big[
(-4+a_{2})a_{2}+16(-2+b_{1}^{2})
\Big]b_{2}
\nonumber\\
&-
3\Big[
(-16+a_{2})a_{2}+16(-5+b_{1}^{2})
\Big]b_{2}^{2}
-
4a_{1}^{2}
\Big[
144-42a_{3}
+3a_{2}(-2+3a_{2}+4a_{3})
-32b_{1}^{2}
+80b_{1}b_{2}
-33b_{2}^{2}
\Big]
\nonumber\\
&+
2a_{1}
\Big[
-3a_{2}^{3}
+a_{2}^{2}(30-21a_{3})
-16a_{3}(3+b_{1}^{2})
+32(3+9b_{1}^{2}-14b_{1}b_{2}+6b_{2}^{2})
\nonumber\\
&\qquad
+a_{2}(-132a_{3}+48(7+b_{1}^{2})-64b_{1}b_{2}+30b_{2}^{2})
\Big],
\end{align}


\begin{align}
\mathcal{I}
={}&
352 a_{1}^{5}
+75 a_{2}^{5}
-16 a_{1}^{4}(72+25a_{2}-33a_{3})
+a_{2}^{4}(184+225a_{3})
\nonumber\\
&+
32a_{2}^{2}
\Big[
-576-292b_{1}^{2}
+3a_{3}(87+35b_{1}^{2})
+276b_{1}b_{2}
-99b_{2}^{2}
\Big]
\nonumber\\
&+
16a_{2}^{3}
\Big[
-218+147a_{3}+30b_{1}^{2}+58b_{1}b_{2}-21b_{2}^{2}
\Big]
\nonumber\\
&+
256
\Big[
-56(1+4b_{1}^{2}+3b_{1}^{4})
+3a_{3}(7+20b_{1}^{2}+3b_{1}^{4})
+8b_{1}(11+26b_{1}^{2})b_{2}
-15(2+5b_{1}^{2})b_{2}^{2}
\Big]
\nonumber\\
&+
64a_{2}
\Big[
90a_{3}(2+3b_{1}^{2})
-4(113+232b_{1}^{2}+33b_{1}^{4})
+8b_{1}(51+26b_{1}^{2})b_{2}
-3(48+25b_{1}^{2})b_{2}^{2}
\Big]
\nonumber\\
&-
16a_{1}^{3}
\Big[
a_{2}(16+33a_{2}+54a_{3})
-8(-26+3a_{3}-22b_{1}^{2}-14b_{1}b_{2}+6b_{2}^{2})
\Big]
\nonumber\\
&+
8a_{1}^{2}
\Big[
31a_{2}^{3}
+a_{2}^{2}(168-45a_{3})
-16(-64-100b_{1}^{2}+9a_{3}(3+5b_{1}^{2})+12b_{1}b_{2}-9b_{2}^{2})
\nonumber\\
&\qquad
-8a_{2}(-162+63a_{3}+14b_{1}^{2}-30b_{1}b_{2}+9b_{2}^{2})
\Big]
\nonumber\\
&+
2a_{1}
\Big[
167a_{2}^{4}
+12a_{2}^{3}(40+23a_{3})
+64a_{2}
\Big[
32+a_{3}(9+30b_{1}^{2})
-3(8b_{1}-5b_{2})(4b_{1}-3b_{2})
\Big]
\nonumber\\
&\qquad
-64
\Big[
-60-64b_{1}^{2}+132b_{1}^{4}
+6a_{3}(2+5b_{1}^{2})
-8b_{1}(15+26b_{1}^{2})b_{2}
+3(12+25b_{1}^{2})b_{2}^{2}
\Big]
\nonumber\\
&\qquad
+16a_{2}^{2}
(-30+81a_{3}+2(19b_{1}-6b_{2})(b_{1}+3b_{2}))
\Big].
\end{align}

\section{Einstein-Cartan theory}
\label{appendixC}

The purpose of this appendix is to compare the Einstein--Cartan limit of the Poincar\'e gauge theory discussed in the main text with the same calculation written in Hamilton's connection convention. The main text follows Schouten's index convention, whereas this appendix uses Hamilton's convention \cite{HamiltonGR}. The dictionary between the two notations is given at the end of the appendix.

In Hamilton's convention, one has
\begin{align}
    & \nabla_\mu V^\nu=\partial_\mu V^\nu+\Gamma_{\lambda\mu}^\nu V^\lambda \,,\qquad  \nabla_\mu V_\nu=\partial_\mu V_\nu-\Gamma_{\nu\mu}^\lambda V_\lambda\,, \\
    & T_{\mu\nu}^{\sigma} = g^{\sigma \lambda}\Bigl(\Gamma_{\lambda\mu\nu}-\Gamma_{\lambda\nu\mu}\Bigr)\,,\quad \Gamma_{\lambda\mu\nu}\equiv g_{\lambda\sigma }\Gamma^\sigma_{\mu\nu} \,,\\
    & \Gamma_{\lambda\mu\nu} = \overset{\circ}{\Gamma}_{\lambda\mu\nu}+K_{\lambda\mu\nu}\,,\qquad  \overset{\circ}{\Gamma}_{\lambda\mu\nu} = \frac{1}{2}\left(\partial_\mu g_{\lambda\nu}+\partial_\nu g_{\lambda\mu}-\partial_\lambda g_{\mu\nu}\right)\,,\\
    & K_{\mu\nu\lambda}=\frac{1}{2}\Bigl(T_{\mu\nu\lambda}+T_{\nu\lambda\mu}+T_{\lambda\nu\mu}\Bigr)\,,\quad T_{\lambda\mu\nu}\equiv g_{\lambda\sigma }T^\sigma_{\mu\nu}\,,\label{hamilton2}\\
    & K_{\mu\nu \lambda} = - K_{\nu \mu\lambda }\,,\\
    & R_{\sigma \lambda \mu \nu} = \partial_\sigma \Gamma_{\mu\nu\lambda} - \partial_\lambda \Gamma_{\mu \nu\sigma} + \Gamma_{\mu \lambda}^\rho \Gamma_{\rho\nu\sigma} - \Gamma_{\mu\sigma}^\rho \Gamma_{\rho\nu\lambda}\,,\qquad R_{\mu\nu} = g^{\rho\sigma}R_{\rho\mu\sigma\nu}\,.
\end{align}
It follows that
\begin{align}
    R=\overset{\,\circ}{R}+2\overset{\circ}{\nabla}_\alpha K^\alpha
+g^{\lambda\nu}g^{\mu \beta} g^{\sigma \alpha}\Bigl(K_{\alpha\mu\lambda}K_{\sigma \nu \beta}-K_{\alpha\mu\beta}K_{\sigma\nu\lambda}\Bigr)\,,\quad K^\alpha\equiv K^{\alpha\mu}{}_\mu\,.
\end{align}
Therefore, the action can be written as
\begin{align}
    S &= \frac{1}{2\kappa}\int {d^4}x \sqrt{-g} R + S_m\,, \nonumber\\
    & = \frac{1}{2\kappa}\int {d^4}x \sqrt{-g}\left[\overset{\,\circ}{R}
+g^{\lambda\nu}g^{\mu \beta} g^{\sigma \alpha}\Bigl(K_{\alpha\mu\lambda}K_{\sigma \nu \beta}-K_{\alpha\mu\beta}K_{\sigma\nu\lambda}\Bigr)
    \right] + S_m + \text{boundary terms}\,.
\end{align}
The variation with respect to the contorsion gives
\begin{align}
\label{hamilton1}
    K^\beta g^{\alpha\gamma}-K^\alpha g^{\beta \gamma}+K^{\gamma\beta\alpha}-K^{\gamma\alpha\beta}=-2\kappa  \tau^{\alpha\beta\gamma}\,,
\end{align}
where we have defined
\begin{align}
    \tau^{\alpha\beta\gamma}\equiv \frac{1}{\sqrt{-g}} \frac{\delta S_m}{\delta K_{\alpha\beta\gamma}}\,.
\end{align}
Taking the trace of \eqref{hamilton1} gives
\begin{align}
    K^{\beta\gamma\alpha}-K^{\alpha\gamma\beta}=\kappa \left(2 \tau^{\alpha\beta\gamma}+\tau^\beta g^{\alpha\gamma}-\tau^\alpha g^{\beta\gamma}\right)\,,\quad \tau^\alpha \equiv \tau^{\alpha\mu}{}_\mu\,.
\end{align}
Finally, using
\begin{align}
K^{\alpha\gamma\beta}=-K^{\gamma\alpha\beta}=-K^{\gamma\beta\alpha}-T^{\gamma\alpha\beta}=K^{\beta\gamma\alpha}-T^{\gamma\alpha\beta}\,,
\end{align}
we obtain
\begin{align}
    T^{\gamma\alpha\beta}=\kappa\Bigl(2 \tau^{\alpha\beta\gamma}+\tau^\beta g^{\alpha\gamma}-\tau^\alpha g^{\beta\gamma}\Bigr)\,,
\end{align}
and therefore, from \eqref{hamilton2}
\begin{align}
\label{hamilton3}
    K_{\alpha\beta\gamma}= \kappa\Bigl(-\tau_{\alpha\beta\gamma}-\tau_{\alpha\gamma\beta}-\tau_{\gamma\beta\alpha}+\tau_\alpha g_{\beta\gamma}-\tau_\beta g_{\alpha\gamma}\Bigr)\,.
\end{align}
The variation with respect to the metric gives
\begin{align}
    \overset{\,\circ}{R}_{\mu\nu}-\frac{1}{2}\overset{\,\circ}{R} \,g_{\mu\nu}=\kappa T_{\mu\nu}+S_{\mu\nu}\,,\quad T_{\mu\nu}=-\frac{2}{\sqrt{-g}}\frac{\delta S_m}{\delta g^{\mu\nu}}\,,
\end{align}
with 
\begin{align}
    S_{\mu\nu}=K_\mu K_\nu
    +2K^\alpha K_{\alpha(\mu\nu)}
    +2K^{\alpha\beta}{}_{(\mu}K_{\nu)\alpha\beta}
    -K_{(\mu}{}^{\alpha\beta}K_{\nu)\beta\alpha}
    -\frac{1}{2}g_{\mu\nu}\Bigl(K_\alpha K^\alpha-K_{\alpha\beta\gamma} K^{\alpha\gamma\beta}\Bigr)\,.
\end{align}
Finally, substituting \eqref{hamilton3}, we obtain
\begin{align}
    S_{\mu\nu}=&\kappa^2\left(
2\tau_{\mu}\tau_{\nu}
- \tau_{ab\mu}\tau^{ab}{}_{\nu}
+ 4\tau_{ab(\mu}\tau_{\nu)}{}^{ab}
- 2\tau_{\mu}{}^{ab}\tau_{\nu ab} 
- 2\tau_{\mu}{}^{ab}\tau_{\nu ba}
+ 4\tau^{a}\tau_{a (\mu\nu)}\right.\nonumber\\
&\left.+g_{\mu\nu}\Bigl(\frac{1}{2} \tau_{abc}\tau^{abc}+ \tau_{acb}\tau^{abc}- \tau_{a}\tau^{a}\Bigr)\right)\,.
\end{align}
Assuming the Weyssenhoff form $\tau_{\alpha \beta \gamma}=u_\gamma s_{\alpha \beta}$ together with the Frenkel condition $u^\alpha s_{\alpha \beta}=0$, we obtain
\begin{align}
    S_{\mu\nu}=\kappa^2\left(-s_{\alpha\beta}s^{\alpha\beta}\left(u_\mu u_\nu+\frac{1}{2}g_{\mu\nu}\right)+2s_{\mu\alpha}s_\nu{}^\alpha\right)\,.
\end{align}
The result agrees with the Einstein--Cartan limit of the Poincar\'e gauge theory discussed in the main text, namely the special case
\begin{align}
a_1=a_2=a_3=b_1=b_2=0 .
\end{align}
The apparent differences are due only to the different ordering of the connection indices and to the corresponding dictionary between the two notations. More explicitly, the quantities used in the main text, denoted by a superscript \((S)\), and the quantities used in this appendix, denoted by a superscript \((H)\), are related by
\begin{align}
    &\Gamma_{\alpha\beta\gamma}^{(S)}=\Gamma_{\gamma\beta\alpha}^{(H)}\,,\\
    & T_{\alpha\beta\gamma}^{(S)}=-\frac{1}{2}T_{\gamma\alpha\beta}^{(H)}\,,\\
    & K_{\alpha\beta\gamma}^{(S)} = K_{\gamma\beta\alpha}^{(H)}\,,\quad K_{\alpha}^{(S)} = K_{\alpha}^{(H)}\,,\\
    & \tau_{\alpha\beta\gamma}^{(S)}=-\tau_{\gamma\beta\alpha}^{(H)}\,.
\end{align}
Using this dictionary, it is easy to show that the formulas of both conventions are the same.


\section{Canonical and metric energy--momentum tensors}
\label{appendixD}

The purpose of this appendix is to clarify which energy-momentum tensor appears
in the Einstein--Cartan equations when the theory is derived in the vielbein
formalism. In the presence of spin, the variation with respect to the vielbein
defines the canonical energy-momentum tensor \(\Sigma_{\mu\nu}\), which is not
symmetric in general. Its antisymmetric part is fixed by the Lorentz Noether
identity and is tied to the spin current. After eliminating torsion, the
equations can be rewritten by introducing a Belinfante--Rosenfeld improved
tensor \(\Theta_{\mu\nu}\), but this tensor is not identical to the original
canonical source \(\Sigma_{\mu\nu}\). This distinction is important because
different identifications of the matter energy-momentum tensor lead to different
interpretations of the effective spin contribution.

We therefore keep track of the canonical source, the spin current, and the
Belinfante--Rosenfeld improvement explicitly, before specializing the result to
a Weyssenhoff fluid.

The coframe is denoted by \(e^a=e^a{}_\mu dx^\mu\),
with
\begin{align}
    g_{\mu\nu}=\eta_{ab}e^a{}_\mu e^b{}_\nu,
    \qquad
    \eta_{ab}=\mathrm{diag}(-,+,+,+).
\end{align}
The spin connection is metric compatible,
\begin{align}
    \omega_{ab}=-\omega_{ba},
\end{align}
and is related to the spacetime connection by the tetrad postulate (covariant differentiation commute with changes of basis)
\begin{align}
    \partial_\mu e^a{}_\nu+\omega^a{}_{b\mu}e^b{}_\nu
    -\Gamma^\rho{}_{\mu\nu}e^a{}_\rho=0.
\end{align}
The torsion two-form is defined by
\begin{align}
    2T^a=de^a+\omega^a{}_b\wedge e^b=T^a{}_{\mu\nu}dx^\mu\wedge dx^\nu,
\end{align}
The curvature two-form is
\begin{align}
    R^a{}_b=d\omega^a{}_b+\omega^a{}_c\wedge\omega^c{}_b=\frac{1}{2}R_{\mu\nu b}{}^a dx^\mu\wedge dx^\nu,
\end{align}
with spacetime projection
\begin{align}
    R_{\mu\nu\lambda}{}^\rho
    =
    \partial_\mu \Gamma^\rho{}_{\nu\lambda}
    -
    \partial_\nu \Gamma^\rho{}_{\mu\lambda}
    +
    \Gamma^\sigma{}_{\nu\lambda}\Gamma^\rho{}_{\mu\sigma}
    -
    \Gamma^\sigma{}_{\mu\lambda}\Gamma^\rho{}_{\nu\sigma}.
\end{align}
We use the contraction convention
\begin{align}
    R_{\mu\nu}=R_{\rho\mu\nu}{}^\rho,
    \qquad
    R=g^{\mu\nu}R_{\mu\nu}.
\end{align}
The Einstein--Cartan action is
\begin{align}
    S
    =
    \frac{1}{4\kappa}
    \int
    \varepsilon_{abcd}\,
    e^a\wedge e^b\wedge R^{cd}(\omega)
    +
    S_m(e,\omega,\Psi_m),
\end{align}
where \(\kappa=8\pi G\). The matter variations define the canonical
energy-momentum three-form \(\Sigma_a\) and the spin three-form \(s_{ab}\) by
\begin{align}
    \delta S_m
    =
    \int
    \left(
    -\delta e^a\wedge \Sigma_a
    -
    \frac{1}{2}\delta\omega^{ab}\wedge s_{ab}
    \right).
\end{align}
The field equations are
\begin{align}
\label{Eq:EC1}
    & G_a = \kappa \Sigma_a, \\
\label{Eq:EC2}
    & 2\,\bm{\eta}_{abc}\wedge T^c = \kappa s_{ab},
\end{align}
where
\begin{align}
    G_a \equiv \frac{1}{2}\bm{\eta}_{abc}\wedge R^{bc},
    \qquad
    \bm{\eta}_{abc}\equiv *(\theta_a\wedge\theta_b\wedge\theta_c)=\varepsilon_{abcd}e^d\,,
\end{align}
and $\theta_a=\eta_{ab} e^b$ is the coframe basis. These equations (\ref{Eq:EC1},\ref{Eq:EC2}) read in spacetime components
\begin{align}
    & R_{\mu\nu}(\Gamma) - \frac{1}{2}R(\Gamma)g_{\mu\nu} = \kappa \Sigma_{\mu\nu}\,,\\
\label{Eq:EC3}
    & 2T^\sigma{}_{\mu\nu} + 2T_\mu \delta^\sigma_\nu - 2T_\nu \delta^\sigma_\mu = \kappa s_{\mu\nu}{}^\sigma\,,\qquad  T_\mu\equiv T^\sigma{}_{\sigma\mu}\,.
\end{align}
Since \(\Sigma_{\mu\nu}\) is the canonical energy-momentum tensor, it is not
symmetric in general. The antisymmetric part is not an independent equation: it is equivalent to the Lorentz Noether identity obtained from local Lorentz invariance of the matter sector. Taking the trace of Eq. \eqref{Eq:EC3} gives 
\begin{align}
    T_\mu=\frac{\kappa}{4}s_\mu,
\end{align}
and therefore
\begin{align}
     T^\sigma{}_{\mu\nu} = \frac{\kappa}{2} \left( s_{\mu\nu}{}^\sigma + \frac{1}{2}s_\nu\delta^\sigma_\mu - \frac{1}{2}s_\mu\delta^\sigma_\nu \right)\,,\qquad s_\mu\equiv s_{\mu\nu}{}^\nu\,.
\end{align}
We decompose the connection into its Levi-Civita part and contorsion,
\begin{align}
    \Gamma^\rho{}_{\mu\nu}
    =
    \{{}_{\mu\nu}^{\sigma}\}+
    K_{\mu\nu}{}^\rho,
    \qquad
    K_{\mu\nu\rho}=-K_{\mu\rho\nu}.
\end{align}
With our torsion convention,
\begin{align}
    K_{\mu\nu}{}^\sigma-K_{\nu\mu}{}^\sigma
    =
    2T^\sigma{}_{\mu\nu},
\end{align}
and hence
\begin{align}
    K_{\mu\nu\sigma}
    =
    T_{\mu\nu\sigma}
    -
    T_{\nu\sigma\mu}
    +
    T_{\sigma\mu\nu}.
\end{align}
The Ricci tensor decomposes as
\begin{align}
    R_{\mu\nu}(\Gamma)
    =
    \overset{\,*}{R}_{\mu\nu}
    +
    \overset{\,*}{\nabla}_\rho K_{\mu\nu}{}^\rho
    -
    \overset{\,*}{\nabla}_\mu K_{\rho\nu}{}^\rho
    +
    K_{\mu\nu}{}^\sigma K_{\rho\sigma}{}^\rho
    -
    K_{\rho\nu}{}^\sigma K_{\mu\sigma}{}^\rho .
\end{align}
The relation with the spin tensor used in the metric formulation is
\begin{align}
    -\frac{2}{\sqrt{-g}}
    \frac{\delta S_m}{\delta K^{\mu\sigma\rho}}
    =
    s_{\rho\sigma\mu}.
\end{align}
Therefore, if \(\tau_{\mu\sigma\rho}\) denotes the metric spin current, then
\begin{align}
    2\tau_{\mu\sigma\rho}=s_{\rho\sigma\mu}\,.
\end{align}
The torsion-spin relation may then be written as
\begin{align}
    T_{\mu\nu\sigma}
    =
    \frac{\kappa}{2}
    \left(
    -2\tau_{\sigma\mu\nu}
    +
    \tau_\nu g_{\sigma\mu}
    -
    \tau_\mu g_{\sigma\nu}
    \right).
\end{align}
which coincides with the expression obtained in the metric approach.

We now specialize to a Weyssenhoff fluid and take
\begin{align}
    \tau_{\alpha\beta\gamma}
    =
    u_\alpha s_{\beta\gamma},
    \qquad
    s_{\beta\gamma}=-s_{\gamma\beta},
    \qquad
    u^\alpha s_{\alpha\beta}=0.
\end{align}
After eliminating torsion, the symmetric part of the Einstein--Cartan equations
can be written in Levi-Civita form as
\begin{align}
    \overset{\,*}{G}_{\mu\nu}
    =
    \kappa
    \left[
    \Sigma_{(\mu\nu)}
    -
    \overset{\,*}{\nabla}_\rho
    \left(
    u_\mu s_\nu{}^\rho
    +
    u_\nu s_\mu{}^\rho
    \right)
    \right]
    -
    \kappa^2
    s_{\alpha\beta}s^{\alpha\beta}
    \left(
    u_\mu u_\nu
    +
    \frac{1}{2}g_{\mu\nu}
    \right),
\end{align}
while the antisymmetric part gives
\begin{align}
    \overset{\,*}{\nabla}_\rho
    \left(
    u^\rho s_{\mu\nu}
    \right)
    =
    -\Sigma_{[\mu\nu]}.
\end{align}
Introducing the Belinfante--Rosenfeld improved tensor
\begin{align}
    \Theta_{\mu\nu}
    =
    \Sigma_{\mu\nu}
    +
    \overset{\,*}{\nabla}_\rho E^\rho{}_{\mu\nu},
\end{align}
with the standard choice
\begin{align}
    E^\rho{}_{\mu\nu}
    =
    u^\rho s_{\mu\nu}
    +
    u_\nu s^\rho{}_\mu
    +
    u_\mu s^\rho{}_\nu,
\end{align}
one obtains the symmetric effective equation
\begin{align}
    \overset{\,*}{G}_{\mu\nu}
    =
    \kappa \Theta_{\mu\nu}
    -
    \kappa^2
    s_{\alpha\beta}s^{\alpha\beta}
    \left(
    u_\mu u_\nu
    +
    \frac{1}{2}g_{\mu\nu}
    \right).
\end{align}
If the improved tensor is identified with the matter energy-momentum tensor,
\(\Theta_{\mu\nu}\equiv T_{\mu\nu}\), this gives
\begin{align}
    &\overset{\,*}{G}_{\mu\nu}=\kappa T_{\mu\nu}-\kappa^2 s_{\alpha\beta} s^{\alpha\beta}\left(u_\mu u_\nu+\frac{1}{2}g_{\mu\nu}\right)
\end{align}
which corresponds to \cite{Obukhov:1987yu}. This is the origin of the ambiguity in the literature. The tensor
\(\Sigma_{\mu\nu}\) obtained from the vielbein variation is the canonical
energy-momentum tensor and is not symmetric in the presence of spin. After
eliminating torsion, the equations can be rewritten using a Belinfante--Rosenfeld
improved tensor \(\Theta_{\mu\nu}\), which differs from \(\Sigma_{\mu\nu}\) by
spin-current derivative terms. Therefore different identifications of the
matter energy-momentum tensor lead to different interpretations of the effective
spin contribution.

\bibliographystyle{unsrt}
\bibliography{refs} 

\end{document}